\documentclass[preprint,5p]{elsarticle}

\usepackage{amssymb}
\usepackage{amsmath}
\usepackage{hyperref}
\usepackage{xcolor}
\usepackage{makecell}
\usepackage[frozencache]{minted}
\setminted[python]{
    breaklines=true,
    encoding=utf8,
    fontsize=\footnotesize,
    frame=lines
}

\usepackage{placeins} % in order to use floatbarrier

\biboptions{comma,compress}

\newcounter{bla}

\journal{Computer Physics Communications}

\begin{document}

\begin{frontmatter}



\title{{\tt Qibo}: a framework for quantum simulation with hardware acceleration}

\author[label1]{Stavros Efthymiou}
\author[label1,label2]{Sergi Ramos-Calderer}
\author[label3,label2]{Carlos Bravo-Prieto}
\author[label3,label2]{Adri\'an P\'erez-Salinas}
\author[label3,label2,label4]{Diego Garc\'ia-Mart\'in}
\author[label3,label5]{Artur Garcia-Saez}
\author[label1,label6,label2]{Jos\'e Ignacio Latorre}
\author[label7,label1]{Stefano Carrazza\corref{author}}

\cortext[author] {Corresponding author.\\\textit{E-mail address:}
stefano.carrazza@unimi.it\\\textit{Preprint number:} TIF-UNIMI-2020-21}

\address[label1]{Quantum Research Centre, Technology Innovation Institute, Abu Dhabi, UAE.}
\address[label2]{Departament de F\'isica Qu\`antica i Astrof\'isica and Institut de Ci\`encies del Cosmos (ICCUB), Universitat de Barcelona, Barcelona, Spain.}
\address[label3]{Barcelona Supercomputing Center, Barcelona, Spain.}
\address[label4]{Instituto de F\'isica Te\'orica, UAM-CSIC, Madrid, Spain.}
\address[label5]{Qilimanjaro Quantum Tech, Barcelona, Spain.}
\address[label6]{Centre for Quantum Technologies, National University of Singapore, Singapore.}
\address[label7]{TIF Lab, Dipartimento di Fisica, Universit\`a degli Studi di Milano and
INFN Sezione di Milano, Milan, Italy.}

\begin{abstract}
    We present {\tt Qibo}, a new open-source software for fast evaluation of
    quantum circuits and adiabatic evolution which takes full advantage of
    hardware accelerators. The growing interest in quantum computing and the
    recent developments of quantum hardware devices motivates the development of
    new advanced computational tools focused on performance and usage
    simplicity.
    In this work we introduce a new quantum simulation framework that enables
    developers to delegate all complicated aspects of hardware or platform
    implementation to the library so they can focus on the problem and quantum
    algorithms at hand.
    This software is designed from scratch with simulation performance, code
    simplicity and user friendly interface as target goals. It takes advantage
    of hardware acceleration such as multi-threading CPU, single GPU and
    multi-GPU devices.
\end{abstract}


\end{frontmatter}

\noindent
{\bf PROGRAM SUMMARY}
\\

\begin{small}
\noindent
{\em Program Title:} {\tt Qibo} \\
\\
{\em Program URL:} \url{https://github.com/qiboteam/qibo}\\
\\
{\em Licensing provisions:} Apache-2.0 \\
\\
{\em Programming language:} {\tt Python, C/C++} \\
\\
{\em Nature of the problem:} Simulation of quantum circuits and adiabatic evolution
requires implementation of efficient linear algebra operations. The simulation
cost in terms of memory and computing time is exponential with the number of
qubits.\\
\\
{\em Solution method:} Implementation of algorithms for the simulation of
quantum circuits and adiabatic evolution using the dataflow graph infrastructure
provided by the TensorFlow framework in combination with custom operators.
Extension of the algorithm to take advantage of multi-threading CPU, single GPU
and multi-GPU setups.\\

\end{small}

%\tableofcontents

\section{Introduction and motivation}
\label{sec:introduction}

During the last decade, we have observed an impressive fast development of
quantum computing hardware. Nowadays, quantum processing units (QPUs) are based
on two approaches, the quantum circuit and quantum logic gate-based model
processors as implemented by Google~\cite{google}, IBM~\cite{ibmq},
Rigetti~\cite{rigetti} or Intel~\cite{intel}, and the annealing quantum
processors such as D-Wave~\cite{dwave,dwaveneal}. The development of these devices and the
achievement of quantum advantage~\cite{48651} are clear indicators that a
technological revolution in computing will occur in the coming years.

The quantum computing paradigm is based on the hardware implementation of
qubits, the quantum analogue to bits, which are used as the representation of
quantum states. Currently, quantum computer manufacturers provide systems
containing up to dozens of qubits for circuit-based quantum processors, while
annealing quantum processors can reach thousands of qubits.
Thanks to the qubits representation it is possible to implement quantum
algorithms based on different approaches such as the quantum Fourier
transform~\cite{QFT}, amplitude amplification and estimation~\cite{Brassard},
search for elements in unstructured databases~\cite{grover1996fast,Grover_1998},
BQP-complete problems~\cite{nielsen2001quantum}, and hybrid quantum-classical
models~\cite{Moll_2018}. These algorithms are the possible key solution for
different types of problems such as optimization~\cite{qaoa} and prime
factorization~\cite{shor}. However, in several cases, an algorithm's
implementation may require systems with large number of qubits, thus even if in
principle we can simulate the behaviour of quantum hardware devices using
quantum mechanics on classical computers, the computational performance becomes
quickly unpractical due to the exponential scaling of memory and time.

The quantum computer simulation on classical hardware is still quite relevant in
the current research stage, because thanks to simulation, researchers can
prototype and study a priori the behaviour of new algorithms on quantum
hardware. In terms of simulation techniques, there are at least three common
approaches such as the linear algebra implementation of the quantum-mechanical
wave-function propagation, the Feynman path-integral formulation of quantum
mechanics~\cite{boixo2017simulation,chen2018classical} and tensor
networks~\cite{tensornetworks}.

The simulation of circuit-based quantum processors is already implemented by
several research collaborations and companies. Some notable examples of
simulation software which are based on linear algebra approach are {\tt
Cirq}~\cite{cirq} and TensorFlow Quantum ({\tt TFQ})~\cite{tfq} from Google,
{\tt Qiskit} from IBM Q~\cite{Qiskit}, {\tt PyQuil} from Rigetti~\cite{pyquil},
{\tt Intel-QS} ({\tt qHipster}) from Intel~\cite{intelqs} , {\tt
QCGPU}~\cite{qcgpu} and {\tt Qulacs}~\cite{Qulacs}, among
others~\cite{Jones_2019,10.1007/978-3-319-27119-4_17,Steiger_2018,qsharp,zulehner2017advanced,pednault2017paretoefficient,PhysRevLett.116.250501,DERAEDT2007121,Fried_2018,Villalonga_2019,luo2019yaojl,bergholm2018pennylane,10.1145/3310273.3323053,10.1007/978-3-030-50433-5_35,Jones_2020,Chen_2018,EasyChair:4050,meyerov2020simulating,moueddene2020realistic,wang2020quantum}.
While the simulation techniques and hardware-specific configurations are well
defined for each simulation software, despite the availability of recent
implementations based on Field Programmable Gate Arrays
(FPGAs)~\cite{10.1007/s10825-018-1287-5,doi:10.1021/acs.jctc.9b01284}, there are
no simulation tools that can take full advantage of hardware acceleration in
single and double precision computations, through a simple interface which
allows the user to switch from multithreading CPU, single GPU, and distributed
multi-GPU/CPU setups.
On the other hand, from the point of view of quantum annealing computation, in
particular adiabatic quantum computation, there are several examples of
applications in the
literature~\cite{farhi2000quantum,Kadowaki_1998,Crosson_2016}. However,
classical simulation of adiabatic evolution algorithms used by this
computational paradigm are not systematically implemented in public libraries.

\begin{table}[t]
    \centering
    \begin{tabular}{|l|l|}
    \hline
    Module & Description \\ \hline \hline
    {\tt models} & \makecell[l]{{\tt Qibo} models (details in Table~\ref{tab:qibomodels}).} \\ \hline
    {\tt gates} & \makecell[l]{Quantum gates that can be \\ added to {\tt Qibo} circuit.} \\ \hline
    {\tt callbacks} & \makecell[l]{Calculation of physical quantities \\ during circuit simulation.} \\ \hline
    {\tt hamiltonians} & \makecell[l]{Hamiltonian objects supporting \\ matrix operations and \\ Trotter decomposition.} \\ \hline
    {\tt solvers} & \makecell[l]{Integration methods used for \\ time evolution.} \\ \hline
    \end{tabular}
    \caption{Modules supported by {\tt Qibo 0.1.0}.}
    \label{tab:qibomodules}
\end{table}
\begin{table}[t]
    \centering
    \begin{tabular}{|l|l|}
    \hline
    {\tt Qibo} model & Description \\ \hline \hline
    {\tt Circuit} & \makecell[l]{Basic circuit model containing \\ gates and/or measurements.} \\ \hline
    {\tt DistributedCircuit} & \makecell[l]{Circuit that can be executed \\ on multiple devices.} \\ \hline
    {\tt QFT} & \makecell[l]{Circuit implementing the \\ Quantum Fourier Transform.} \\ \hline
    {\tt VQE} & \makecell[l]{Variational Quantum \\ Eigensolver.\\  Supports optimization of the \\ variational parameters.}  \\ \hline
    {\tt QAOA} & \makecell[l]{Quantum Approximate \\ Optimization Algorithm. \\  Supports optimization of the \\ variational parameters.}  \\ \hline
    {\tt StateEvolution} & \makecell[l]{Unitary time evolution of \\ quantum states under a \\ Hamiltonian.} \\ \hline
    {\tt AdiabaticEvolution} & \makecell[l]{Adiabatic time evolution \\ of quantum states. \\ Supports optimization of \\ the scheduling function.} \\\hline
    \end{tabular}
    \caption{{\tt qibo.models} implemented in {\tt Qibo 0.1.0}.}
    \label{tab:qibomodels}
\end{table}

In this work, we present the {\tt Qibo}
framework~\cite{stavros_efthymiou_2020_3997195} for quantum simulation with
hardware acceleration (code available at~\cite{qibo_github}). {\tt Qibo} is
designed with three target goals: a simple application programming interface
(API) for quantum circuit design and adiabatic quantum computation, a
high-performance simulation engine based on hardware acceleration tools, with
particular emphasis on multithreading CPU, single GPU and multi-GPU setups, and
finally, a clean design pattern to include classical/quantum hybrid algorithms.
In general the inclusion of hardware acceleration support requires a good
knowledge of multiple programming languages such as {\tt C/C++} and {\tt
Python}, and hardware specific frameworks such as {\tt CUDA}~\cite{CUDA}, {\tt
OpenCL}~\cite{OPENCL} and {\tt OpenMP}~\cite{openmp}. However, given that the
knowledge of each of these tools could be a strong technical barrier for users
interested in custom circuit designs, and subsequently, the simulation of new
quantum and hybrid algorithms, {\tt Qibo} proposes a framework build on top of
the TensorFlow~\cite{tensorflow2015:whitepaper} library which reduces the effort
required by the user. Tables~\ref{tab:qibomodules} and~\ref{tab:qibomodels}
provide an overview of the basic modules and models that are implemented in the
current version of {\tt Qibo 0.1.0}.

The {\tt Qibo} framework will become the entry point in terms of API and
simulation engine for the middleware of a new quantum experimental research
collaboration coordinated by~\cite{TII,QQT}.

The paper is organized as follows. In Sec.~\ref{sec:methodology} we present
the technical aspects of the {\tt Qibo} framework, highlighting the code
structure, algorithms, and features. In Sec.~\ref{sec:benchmarks}, we show
benchmarking results comparing the {\tt Qibo} simulation performance with other
popular libraries. The Sec.~\ref{sec:applications} is dedicated to
applications provided by {\tt Qibo} as examples. Finally, in
Sec.~\ref{sec:outlook} we present our conclusion and future development
direction.

\section{Technical implementation}
\label{sec:methodology}

In this section we present the technical structure of {\tt Qibo}, an open-source
library for quantum circuit definition and simulation which takes advantage of
hardware accelerators such as GPUs.

\subsection{Acceleration paradigm}

Hardware acceleration combines the flexibility of general-purpose processors, such as CPUs, with the efficiency of fully customized hardware, such as GPUs, increasing efficiency by orders of magnitude.

In particular, hardware accelerators such as GPUs with a large number of cores and memory are getting popular thanks to their great efficiency in deep learning applications. Open-source frameworks such as TensorFlow simplify the development strategy by reducing the required hardware knowledge from the developer's point of view.

In this context, {\tt Qibo} implements quantum circuit simulation using
TensorFlow primitives and custom operators together with job scheduling for
multi-GPU synchronization. The choice of TensorFlow as the backend development
framework for {\tt Qibo} is motivated by its simple mechanism to write efficient
{\tt Python} code which can be distributed to hardware accelerators without
complicated installation procedures.

\subsection{Code structure}

\begin{figure}
    \centering
    \includegraphics[width=0.45\textwidth]{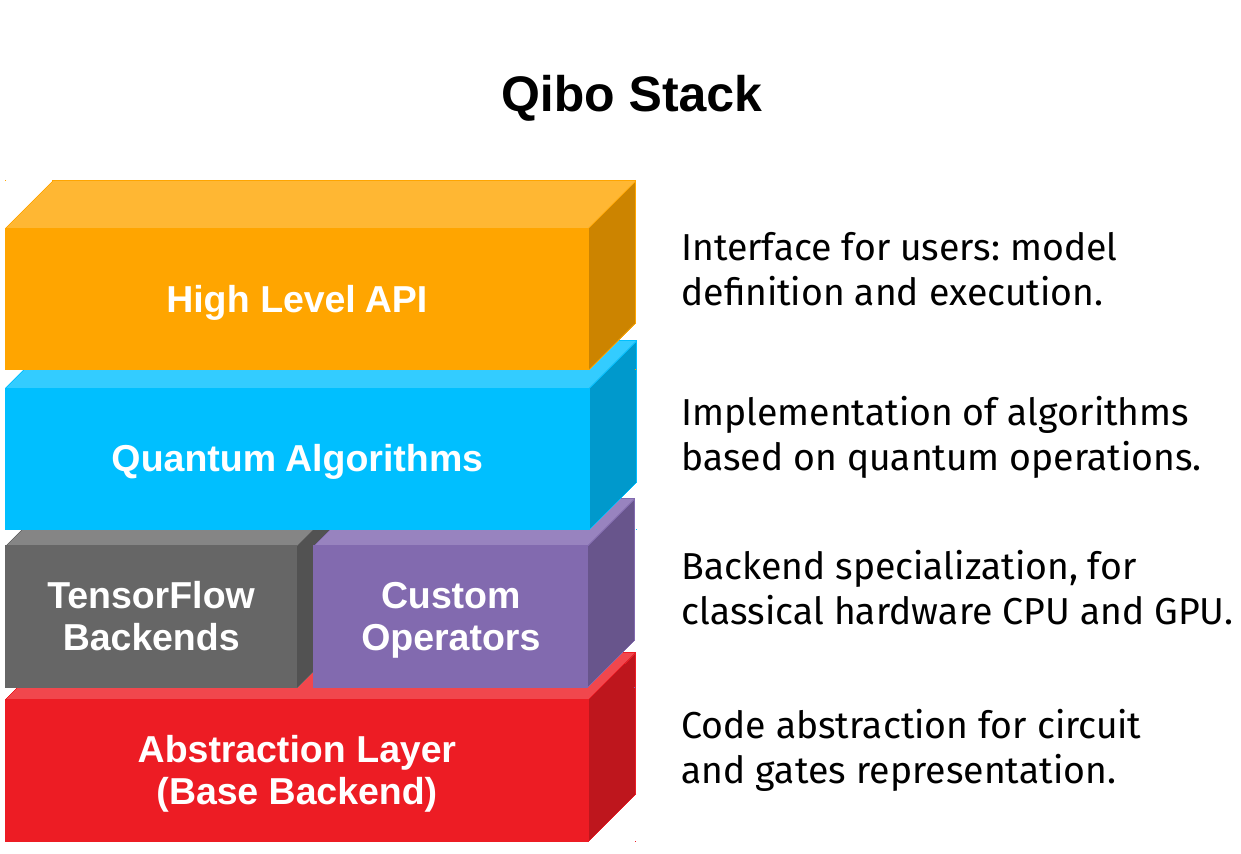}
    \caption{Schematic view of the {\tt Qibo} structure design.}
    \label{fig:stack}
\end{figure}

In Fig.~\ref{fig:stack} we show a schematic representation of the code
structure. The ground layer represents the base abstraction layer, where the circuit structure and gates are defined.
On top of the abstraction layer, we specialize the simulation system using TensorFlow and {\tt numpy}~\cite{numpybook} primitives.
The backend layers are required in order to build quantum algorithms such as the
Variation Quantum Eigensolver (VQE)~\cite{vqe}, perform measurement shots, etc.
These algorithms are implemented in such a way that there is no direct dependency
on the backend specialization. Furthermore, several models delivered by {\tt Qibo}, such
as VQE, QAOA and adiabatic evolution, require minimization techniques provided by
external libraries, in particular TensorFlow for stochastic gradient descent,
Scipy~\cite{2020SciPy-NMeth} for quasi-Newton methods and
CMA-ES~\cite{hansen2006cma} for evolutionary optimization.
Finally, we provide the entry point for code usage through a simple high-level API in {\tt Python}.

\subsection{Backends and algorithms}

{\tt Qibo} simulates the behavior of quantum circuits using dense complex state
vectors $\psi (\sigma _1, \sigma _2,\dots , \sigma _N) \in \mathbb{C}$ in the
computational basis where $\sigma _i \in \{ 0, 1\}$ and $N$ is the total number
of qubits in the circuit. The main usage scheme is the following:
\begin{minted}{python}
import numpy as np
from qibo import models, gates

# create a circuit for N=3 qubits
circuit = models.Circuit(3)

# add some gates in the circuit
circuit.add([gates.H(0), gates.X(1)])
circuit.add(gates.RX(0, theta=np.pi/6))

# execute the circuit and obtain the
# final state as a tf.Tensor
final_state = circuit()
\end{minted}
where {\tt qibo.models.Circuit} is the core {\tt Qibo} object and holds a queue of
quantum gates. Each gate corresponds to a matrix
$G(\boldsymbol{\tau}, \boldsymbol{\tau'}) = G(\tau_1,\dots ,\tau_{N_\mathrm{targets}}, \tau_1',\dots ,\tau_{N_\mathrm{targets}}')$
and acts on the state vector via the matrix multiplication
\begin{equation}\label{eq:gatemul}
    \psi'(\sigma_1, \ldots, \sigma_N) = \sum _{\boldsymbol{\tau'}} G(\boldsymbol{\tau}, \boldsymbol{\tau'})\psi(\sigma_1,\ldots,\boldsymbol{\tau'},\ldots,\sigma_N)
\end{equation}
where the sum runs over qubits targeted by the gate.

When a circuit is executed, the state vector is transformed by applying matrix
multiplications for every gate in the queue. By default, the result of circuit
execution is the final state vector $\psi'$ after all
gates have been applied. If measurement gates are used then the returned result
contains measurement samples following the distribution corresponding to the
final state vector. The computational difficulty in this calculation is that the
dimension of $\psi$ increases exponentially with the
number of qubits $N$ in the circuit.

{\tt Qibo} provides three different backends for implementing the matrix
multiplication of Eq.~(\ref{eq:gatemul}), all based on TensorFlow 2.
There are two backends ({\tt defaulteinsum} and {\tt matmuleinsum}) based on
TensorFlow native operations and the {\tt custom} backend that uses custom
{\tt C++} operators. Table~\ref{tab:backendfeatures} provides a feature
comparison between the different backends. The default backend is {\tt custom},
however the user can easily switch to a different backend using the
{\tt qibo.set\_backend()} method.

In the two TensorFlow backends the state vector $\psi$ is
stored as a rank-$N$ TensorFlow tensor ({\tt tf.Tensor}) and gate matrices as
rank-$2N_\mathrm{targets}$ tensors. In the {\tt defaultein-\\sum} backend, matrix
multiplication is implemented using the {\tt tf.einsum} method, while in the
{\tt matmuleinsum} backend using {\tt tf.matmul}. In the latter case, the state
has to be transposed and reshaped to $(2^{N_\mathrm{targets}}, 2^{N -
N_\mathrm{targets}})$ shape using the {\tt tf.transpose} and {\tt tf.reshape}
operations, as {\tt tf.matmul} by definition supports only matrix (rank-2
tensor) multiplication. The motivation to have both implementations is justified
by performance: the {\tt defaulteinsum} backend is faster on GPUs while the {\tt
matmuleinsum} is more efficient on CPU. The main advantage of using backends
based on native TensorFlow primitives is that support for backpropagation is
inherited automatically. This may be useful when gradient-descent-based
minimization schemes are used to optimize variational quantum circuits. On the
other hand, TensorFlow operations create multiple state vector copies, increasing execution time and memory usage, particularly for large qubit numbers.

In order to increase simulation performance and reduce memory usage, we
implemented custom TensorFlow operators that perform Eq.~(\ref{eq:gatemul}). The state is stored as a vector with $2^N$ components, and the indices of its components that should be updated during the matrix multiplication are calculated by the custom operators during each gate application. This allows all
gate applications to happen in-place without requiring any copies of the state
vector, thus reducing memory requirements to the minimum ($2^N$ complex numbers
for $N$ qubits). Furthermore, the sparsity of some common quantum gates is
exploited to increase performance. For example, the $Z$ gate is applied by
flipping the sign of only half of the state components while the rest remain
unaffected. Custom operators are coded using {\tt C++} and support
multi-threading via TensorFlow's thread pool implementation. An additional {\tt
CUDA} implementation is provided for all operators to allow execution on GPU.

\begin{table}
    \begin{tabular}{|l|c|c|}
    \hline
     & Native & Custom \\ \hline \hline
    Backend names & \makecell{{\tt defaulteinsum} \\ {\tt matmuleinsum}} & {\tt custom} \\ \hline
    GPU support & \checkmark & \checkmark \\ \hline
    Distributed computation &  & \checkmark \\ \hline
    In-place state updates &  & \checkmark \\ \hline
    Measurements & \checkmark & \checkmark \\ \hline
    Controlled gates & \checkmark & \checkmark \\ \hline
    Density matrices / noise & \checkmark &  \\ \hline
    Callbacks & \checkmark & \checkmark \\ \hline
    Gate fusion & \checkmark & \checkmark \\ \hline
    Backpropagation & \checkmark &  \\ \hline
    \end{tabular}
    \caption{Features support for each calculation backend.}
    \label{tab:backendfeatures}
\end{table}

\subsection{Circuit simulation features}\label{sec:features}

{\tt Qibo} provides several features aiming to make the simulation of quantum
circuits for research purposes easier. In this section, we describe some of these
features.

\subsubsection{Controlled gates}

All {\tt Qibo} gates can be controlled by an arbitrary number of qubits.
Both in the native TensorFlow and custom backends, these gates are applied
using the proper indexing of the state vector, avoiding the creation of
large gate matrices.

\subsubsection{Measurements}

{\tt Qibo}'s measuring mechanism works by sampling the final state vector
once all gates of a circuit are applied. Sampling is handled by the
{\tt tf.random.categorical} method. A flexible measurement API is provided, which allows the user to view measurement results in binary or decimal format
and as raw measurement samples or frequency dictionaries.
It is also possible to group multiple qubits in the same register and
perform collective measurements.
Note that no density matrix is computed at this step. The final result is a
trace-out of the outcome probability in unmeasured qubits.

\subsubsection{Density matrices and noise}

Native TensorFlow backends can simulate density matrices in addition to
state vectors. This allows the simulation of noisy circuits using the
channels that are provided as {\tt qibo.gates}. By default {\tt Qibo} uses
state vectors for simulation. However, it switches automatically to density
matrices if a channel is found in the circuit or if the user uses a density
matrix as the initial state of a circuit execution. Density matrices are not
yet implemented in the {\tt custom} backend but will be
included in future releases.

\subsubsection{Callbacks}

The callback functions allow the user to perform calculations on intermediate state vectors during
a circuit execution. A callback example which is implemented in {\tt Qibo} is entanglement entropy. This allows the user to track how entanglement
changes as the state is propagated through the circuit's gates. Other callbacks
implemented in {\tt Qibo} include the {\tt callbacks.Energy} which calculates
the energy (expectation value of a Hamiltonian) of a state or
{\tt callbacks.Gap} which calculates the gap of the adiabatic evolution
Hamiltonian (we refer to Sec.~\ref{sec:timeevolution} for more details).

\subsubsection{Gate fusion}

In some cases, particularly for large qubit numbers, it is more efficient to
fuse several gates by multiplying their respective unitary matrices and
multiply the resulting matrix to the state vector, instead of applying the
original gates one-by-one. {\tt Qibo} provides a simple method
({\tt circuit.fuse()}) to fuse circuit gates up to a two-qubit $4\times 4$
matrix. Additionally, the {\tt VariationalLayer} gate is provided for
efficient simulation of variational circuits that consist of alternating
layers between one-qubit rotations and two-qubit entangling gates.
For more details on this we refer to Sec.~\ref{sec:variationalcircuit}.

\subsection{Distributed computation}\label{sec:distributed}

{\tt Qibo} allows execution of circuits on multiple devices with focus on
systems with multiple GPU configurations. As demonstrated in Sec.~\ref{sec:benchmarks},
GPUs are much faster than a typical CPU for circuit simulation, however, they are
limited by their internal memory. A typical high-end GPU nowadays has 12-16GB of
memory, allowing simulation of up to 29 qubits (30 qubits with single precision
numbers) using {\tt Qibo}'s {\tt custom} backend. For larger qubits, the user has
to use a CPU with sufficient random-access memory (RAM) size or rely on
distributed configurations.

{\tt Qibo} provides a simple API to distribute large circuits to multiple devices.
For example, a circuit can be executed on multiple GPUs by passing an
{\tt accelerators} dictionary when creating the corresponding
{\tt qibo.models.Circuit} object, as:
\begin{minted}{python}
from qibo.models import Circuit
circuit = Circuit(
        nqubits=30,
        accelerators={"/GPU:0": 1, "/GPU:1": 1}
        )
\end{minted}
Dictionary keys define which devices will be used and the values the number of
times each device will be used. Note that a single device can be used more than
once to increase the number of ``logical'' devices. For example, a single GPU can
be reused multiple times to exceed the limit of 29 qubits, making the distributed
implementation useful even for systems with a single GPU.

Device re-usability is allowed by exploiting the system's RAM. The full state
vector is stored in RAM while parts of it are transferred to the available GPUs
to perform the matrix multiplications. This state partition to pieces is inspired
by techniques used in multi-node quantum circuit simulation~\cite{distributedlarose, qhipster}.
More specifically, if $N_\mathrm{devices}$ are available, the state is partitioned
by selecting $\log _2 N_\mathrm{devices}$ qubits (called \textit{global} qubits~\cite{petabytesim})
and indexing according to their binary values. For example, if $N_\mathrm{devices}=2$
and the first qubit is selected as the global qubit, the state of size $2^N$ is
partitioned to two pieces $\psi (0, \sigma _2,\dots ,\sigma _N)$ and
$\psi (1, \sigma _2,\dots ,\sigma _N)$ of size $2^{N - 1}$. Gates targeting
local (non-global) qubits are directly applied by performing the corresponding
matrix multiplication on all logical devices. Gates targeting global qubits cannot
be applied without communication between devices. The scheme that we currently
follow to apply such gates is to move their targets to local qubits by adding
SWAP gates between global and local qubits. These SWAP gates are applied on CPU,
where the full state vector is available. All gates between SWAPs target local
qubits and are grouped and applied together to minimize the CPU-GPU communication.

If logical devices correspond to distinct physical devices, the matrix
multiplications are parallelized among physical devices using {\tt
joblib}~\cite{joblib}.

In terms of memory, the distributed implementation described above is restricted
only by the total amount of RAM available for the system's CPU and not the GPU
memory. The main bottleneck is related to CPU-GPU communication, and therefore the performance depends on the number of SWAP gates required to move all
gates' targets to local qubits. This number depends on the circuit structure. As
presented in the practical examples of Sec.~\ref{sec:benchmarks}, a multi-GPU
configuration can provide significant speed-up compared to using only the CPU.

\subsection{Time evolution}\label{sec:timeevolution}
In addition to the circuit simulation presented in the previous sections,
{\tt Qibo} can be used to simulate a unitary time evolution of quantum states.
Given an initial state vector $\left | \psi _0 \right \rangle $ and an evolution
Hamiltonian $H$, the goal is to find the state $\left | \psi (T) \right \rangle $
after time $T$, so that the  time-dependent Schr\"odinger equation
\begin{equation}
    i\partial _t \left | \psi (t) \right \rangle = H\left | \psi (t) \right \rangle
\end{equation}
is satisfied. Note that the Hamiltonian may have explicit time dependence.

An application of time evolution relevant to quantum computation is the
simulation of adiabatic quantum computation~\cite{farhi2000quantum}.
In this case the evolution Hamiltonian takes the form
\begin{equation}
    H(t) = (1 - s(t)) H_0 + s(t) H_1
\end{equation}
where $H_0$ is a Hamiltonian whose ground state is easy to prepare and is used
as the initial condition, $H_1$ is a Hamiltonian whose ground state is hard to
prepare and $s(t)$ is a scheduling function.
According to the adiabatic theorem, for proper choice of $s(t)$ and
total evolution time $T$, the final state $\left | \psi (T) \right \rangle $
will approximate the ground state of the ``hard'' Hamiltonian $H_1$.

The code below shows how {\tt Qibo} can be used to simulate adiabatic
evolution for the case where the "hard" Hamiltonian is the critical transverse
field Ising model, mathematically:
\begin{equation}\label{eq:tfimhamiltonian_init}
H_0 = -\sum _{i=0}^NX_i
\end{equation}
and
\begin{equation}\label{eq:tfimhamiltonian}
    H_1 = -\sum _{i=0}^N(Z_iZ_{i + 1} + hX_i)
\end{equation}
where $X_i$ and $Z_i$ represent the matrices acting on the $i$-th qubit
and $h=1$.

\begin{minted}{python}
from qibo import models, hamiltonians

# define Hamiltonian objects for 4 qubits
h0 = hamiltonians.X(4)
h1 = hamiltonians.TFIM(4, h=1.0)

# define evolution model with linear scheduling
s = lambda t: t
evolve = models.AdiabaticEvolution(h0, h1, s, dt=1e-2)

# execute the model for T=3 and obtain
# the final state as a tf.Tensor
final_state = evolve(final_time=3)
\end{minted}

Note that a list of callbacks may be passed to the definition of
the {\tt models.AdiabaticEvolution} object, which allows the user to
track various quantities during the evolution.

In terms of implementation, {\tt Qibo} uses two different methods to simulate
time evolution. The first method requires constructing the full $2^N\times 2^N$
matrix of $H$ and uses an ordinary differential equation (ODE) solver to
integrate the time-dependent Schr\"ondiger equation.
The default solver ({\tt "exp"}) is using standard matrix exponentiation to
calculate the evolution operator $e^{-iH\delta t}$ for a single time step
$\delta t$ and applies it to the state vector via the matrix multiplication
\begin{equation}
    \left | \psi (t + \delta t) \right \rangle = e^{-iH\delta t} \left | \psi (t) \right \rangle
\end{equation}
This is repeated until the specified final time $T$ is reached. In addition,
{\tt Qibo} provides two Runge-Kutta
integrators~\cite{runge_c_1895_2178704,kutta}, of fourth-order ({\tt "rk4"}) and
fifth-order ({\tt "rk45"}). Using one of these these solvers the matrix
exponentiation step however such approach is less accurate when compared to the
default {\tt "exp"} solver. The operations used in this method are based on
TensorFlow primitives and particularly the {\tt tf.matmul} and {\tt
tf.linalg.expm}.

The second time evolution method is based on the Trotter decomposition, as
presented in Sec.~4.1 of~\cite{Paeckel:2019yjf}. For local Hamiltonians that
contain up to $k$-body interactions, the evolution operator $e^{-iH\delta t}$
can be decomposed to $2^k \times 2^k$ unitary matrices and therefore time
evolution can be mapped to a quantum circuit consisting of $k$-qubit gates.
{\tt Qibo} provides an additional Hamiltonian object
({\tt qibo.hamiltonians.TrotterHamiltonian}) which can be used to generate the
corresponding circuit. Time evolution is then implemented by applying this
circuit to the initial condition. Since time evolution is essentially mapped to
a circuit model, all {\tt Qibo} circuit functionality such as custom operators
(for up to two-qubit gates) and distributed execution (Sec.~\ref{sec:distributed})
may be used with this evolution method.
Furthermore, this allows the direct simulation of an adiabatic evolution by a
circuit-based quantum computer.

\section{Benchmarks}\label{sec:benchmarks}
In this section we benchmark {\tt Qibo} and compare its performance with other
publicly available libraries for quantum circuit simulation. In addition, we
provide results from running circuit simulations on different hardware
configurations supported by {\tt Qibo} and we compare performance between using
single or double complex precision.
Finally, we benchmark {\tt Qibo} for the simulation of adiabatic time evolution, and we compare the performance of different solvers.

The libraries used in the benchmarks are shown in Table~\ref{tab:libraries}. The
default precision and hardware configuration was used for all libraries and was
compared to the equivalent {\tt Qibo} configuration. Single-thread {\tt Qibo}
numbers were obtained using {\tt taskset} utility to restrict the number of
threads because, when running on CPU, {\tt Qibo} utilizes all available threads
by default. For {\tt Qiskit} we have used the default {\tt Qiskit-Aer} simulator.

\begin{table}[t]
\begin{center}
\begin{tabular}{|l|c|c|}
\hline
Library & Precision & Hardware \\ \hline \hline
{\tt Qibo 0.1.0}~\cite{stavros_efthymiou_2020_3997195} & \makecell{single \\ double} & \makecell{multi-thread CPU \\ GPU \\ multi-GPU} \\ \hline
{\tt Cirq 0.8.1}~\cite{cirq} & single & single-thread CPU \\ \hline
{\tt TFQ 0.3.0}~\cite{tfq} & single & single-thread CPU \\ \hline
{\tt Qiskit 0.16.1}~\cite{Qiskit} & double & single-thread CPU \\ \hline
{\tt PyQuil 2.20.0}~\cite{pyquil} & double & single-thread CPU \\ \hline
{\tt IntelQS 2.0.0}~\cite{intelqs} & double & multi-thread CPU \\ \hline
{\tt QCGPU 0.1.1}~\cite{qcgpu} & single & \makecell{multi-thread CPU \\ GPU} \\ \hline
{\tt Qulacs 0.1.10.1}~\cite{Qulacs} & double & \makecell{multi-thread CPU \\ GPU} \\ \hline
\end{tabular}
\end{center}
\caption{Quantum libraries used in the benchmarks with their supported simulation precisions and hardware configurations.}
\label{tab:libraries}
\end{table}

All results presented in this section are produced with an NVIDIA DGX
Station~\cite{dgx}. The machine specification includes 4x NVIDIA Tesla V100 with
32 GB of GPU memory each, and an Intel Xeon E5-2698 v4 with 2.2 GHz
(20-Core/40-Threads) with 256 GB of RAM. The operating system of this machine is
the default Ubuntu 18.04-LTS with CUDA/{\tt nvcc} 10.1, TensorFlow 2.2.0 and
{\tt g++} 7.5. The source code of the benchmark exercise presented in this
section is available in~\cite{stavros_efthymiou_2021_5565343}.

\subsection{Quantum Fourier Transform}\label{sec:qft}
\begin{figure*}[t]
    \centering
    \includegraphics[width=0.45\textwidth]{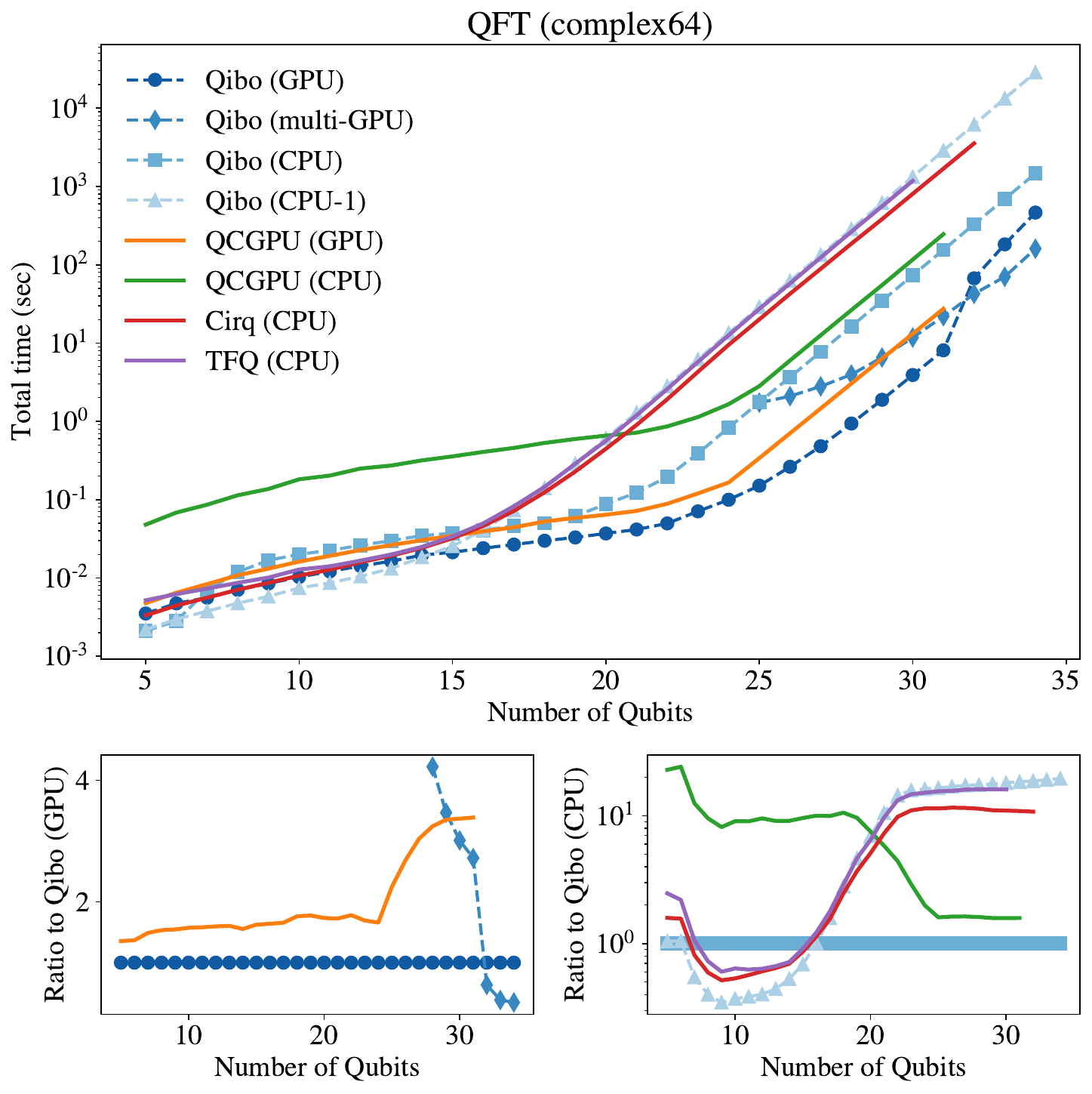}\hspace{1cm}%
    \includegraphics[width=0.45\textwidth]{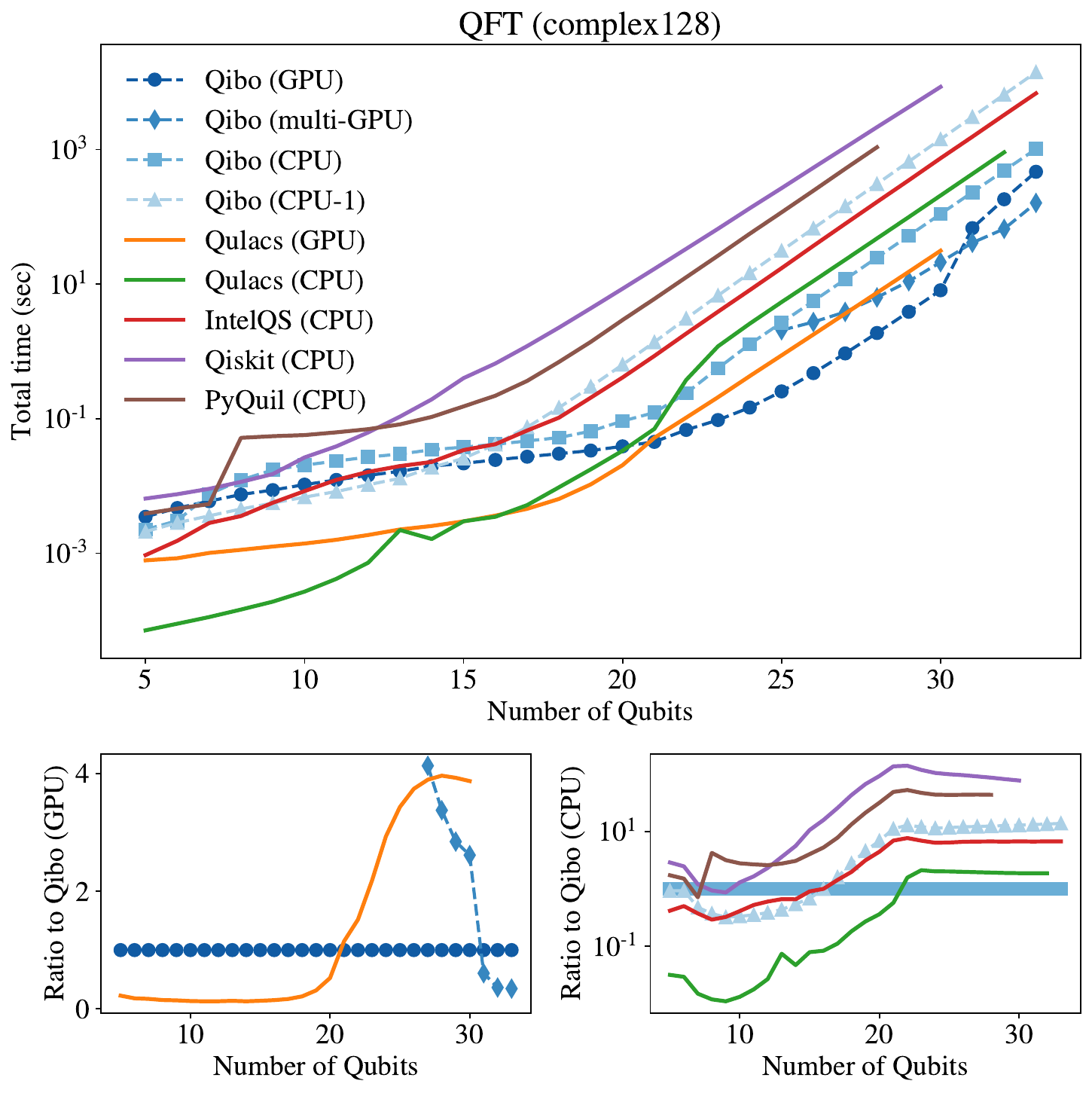}
    \caption{Quantum Fourier Transform simulation performance comparison in
    single precision (left) and double precision (right). Large plots show total
    simulation time as a function of qubit number. Smaller plots show the ratio
    of this time for each library to the corresponding {\tt Qibo} run for GPU
    (left) and CPU (right).}
    \label{fig:qft}
\end{figure*}

The first circuit we used for benchmarks is the Quantum Fourier Transform
(QFT)~\cite{QFT}. This circuit is used as a subroutine in many quantum
algorithms and thus constitutes an example with great practical importance.
The gates used in this circuit are H, CZPow, and SWAP, all of which are available
in {\tt Qibo} and other used libraries, except {\tt QCGPU} where SWAP was implemented
using three CNOT gates.

Results for the QFT circuit are shown in Fig.~\ref{fig:qft}. It is natural to
discuss two regimes separately, the small circuit regime consisting of up to 20
qubits and the large circuit regime for more than 20 qubits. We observe that
most libraries offer similar performance in the first regime. {\tt Qulacs} is
the fastest as it is based on compiled {\tt C++} and avoids the {\tt Python}
overhead that exists in all other libraries. Furthermore, we observe that simpler hardware configurations (single-thread CPU) perform better for small circuits than more complex ones (multi-threading or GPU). In the large circuit
regime {\tt Qibo} offers a better scaling than other libraries in both CPU and
GPU. It is also clear that GPU accelerated libraries offer about an order of
magnitude improvement compared to CPU implementations. We note the exact
agreement between {\tt TFQ} and single-thread {\tt Qibo} as both libraries use
TensorFlow as their computation engine.

In terms of memory, {\tt Qibo} can simulate the highest number of qubits
(33 in {\tt complex128} / 34 in {\tt complex64}) possible for the memory
available in the DGX station (256 GB). A single 32 GB GPU can simulate up to 30
qubits (31 in {\tt complex64}), however this number can be extended up to 33
(34 in single precision) using the distributed scheme described in Sec.~\ref{sec:distributed}
to re-use the same GPU device on multiple state vector partitions. The switch
from single to distributed GPU configuration explains the change of scaling in
the last three points of ``{\tt Qibo} (GPU)'' data. The distributed scheme can
achieve an even better scaling if all four available DGX GPUs are used
(``{\tt Qibo} (multi-GPU)'' line). For more details on the performance of
different hardware configurations supported by {\tt Qibo}, we refer
to Sec.~\ref{sec:configurations}.
\begin{figure}[t]
    \centering
    \includegraphics[width=0.25\textwidth]{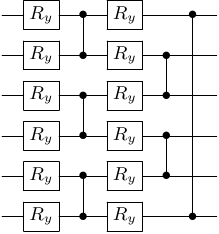}
    \caption{Structure of the variational circuit used in the benchmarks.
    All gates shown in this figure are repeated five times to give the full
    circuit and an additional layer of RY gates is used in the end.}
    \label{fig:varcircuit}
\end{figure}

\subsection{Variational circuit}\label{sec:variationalcircuit}
\begin{figure*}[h!]
    \centering
    \includegraphics[width=0.45\textwidth]{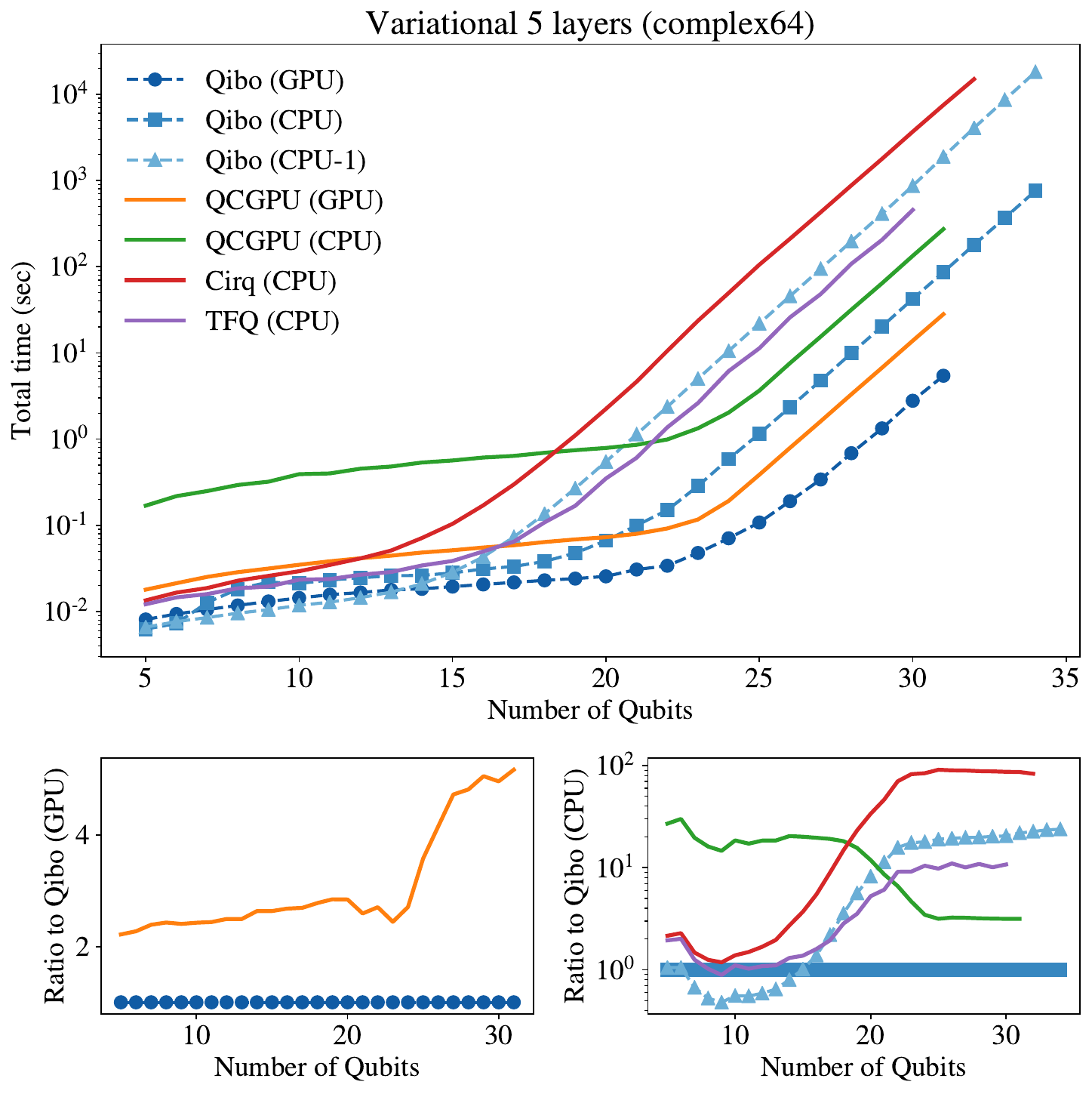}\hspace{1cm}%
    \includegraphics[width=0.45\textwidth]{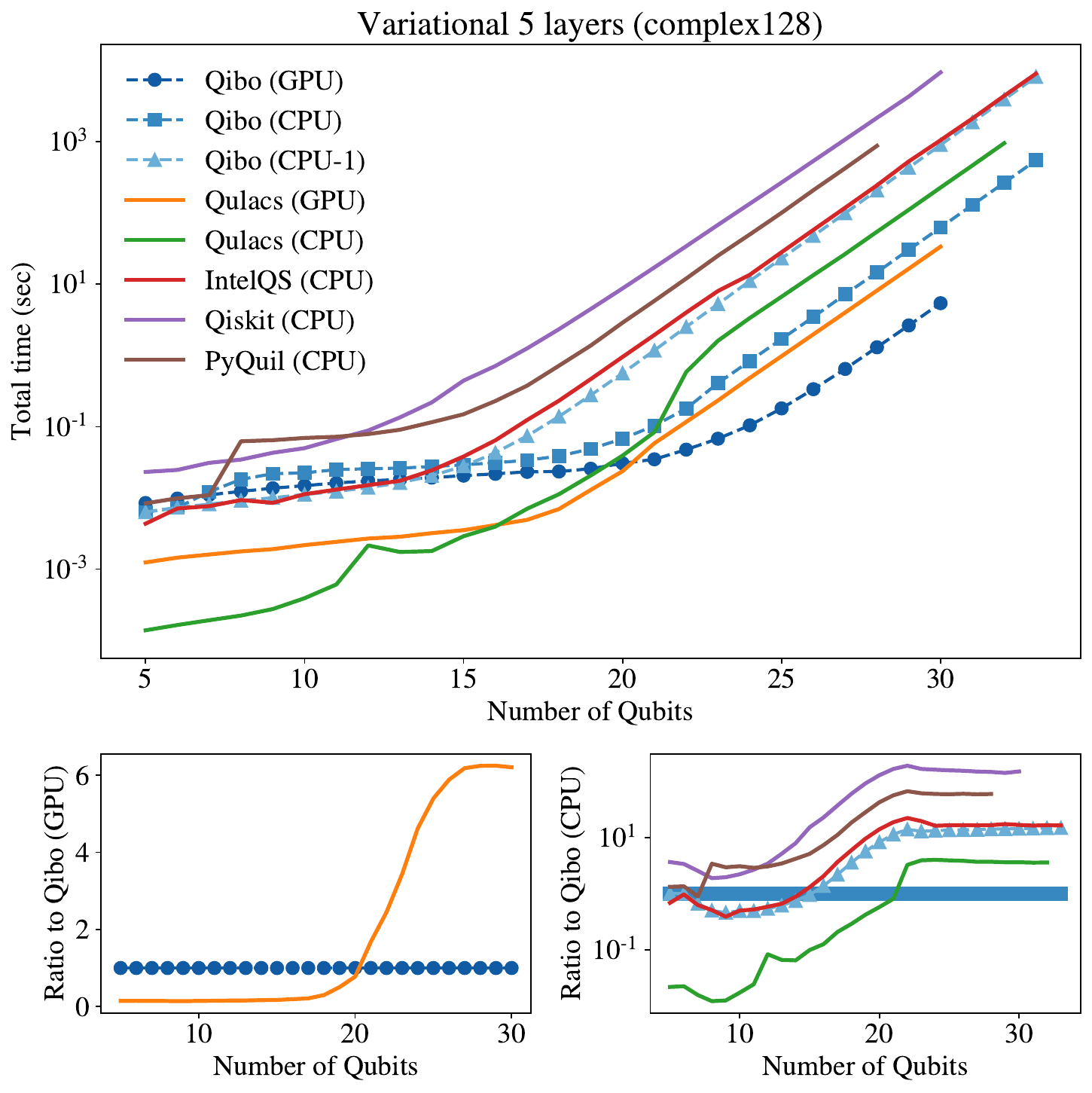}
    \caption{Variational circuit simulation performance comparison in single
    precision (left) and double precision (right). Large plots show total
    simulation time as a function of number of qubits. Smaller plots show the
    ratio of this time for each library to the corresponding {\tt Qibo} run for
    GPU (left) and CPU (right).}
    \label{fig:var5layers}
\end{figure*}
The second circuit used in the benchmarks is inspired by the structure of
variational circuits used in quantum machine learning and similar applications
~\cite{Moll_2018, qaoa}. Such circuits constitute a good candidate
for applications of near-term quantum computers due to their short depth and
are of great interest to the research community. The circuit used in the
benchmark consists of a layer of RY rotations followed by a layer of CZ gates
that entangle neighbouring qubits, as shown in Fig.~\ref{fig:varcircuit}.
The configuration is repeated for five layers and the variational parameters
are chosen randomly from $[0, 2\pi ]$ in all benchmarks.

In Fig.~\ref{fig:var5layers} we plot the results of the variational circuit
benchmark. We observe similar behavior to the QFT benchmarks with all libraries
performing similarly for small qubit numbers and {\tt Qibo} offering superior
scaling for large qubit numbers. The variational circuit is an example where
the gate fusion described in Sec.~\ref{sec:features} is useful. In our
{\tt Qibo} implementation we exploit this by using the {\tt Variational-\\Layer}
gate, which fuses four RY gates with the CZ gate between them and applies them
as a single two-qubit gate. {\tt TFQ} uses a similar fusion algorithm~\cite{tfq}, and unlike the QFT benchmark, it is now noticeably faster than {\tt Cirq}.
All other libraries use the traditional form of the circuit, with each gate
being applied separately.

\begin{figure*}[t]
    \centering
    \includegraphics[width=0.5\textwidth]{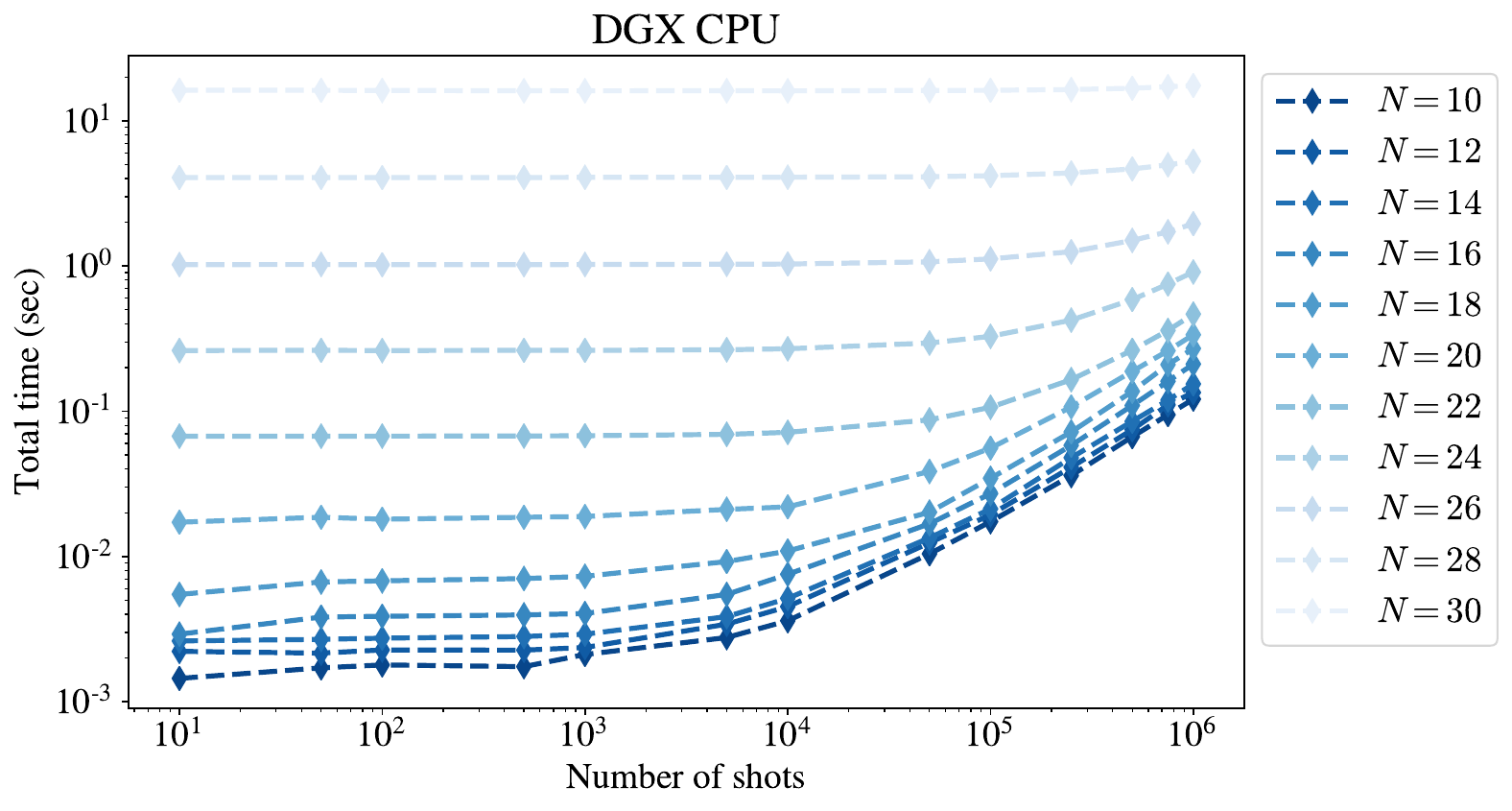}%
    \includegraphics[width=0.5\textwidth]{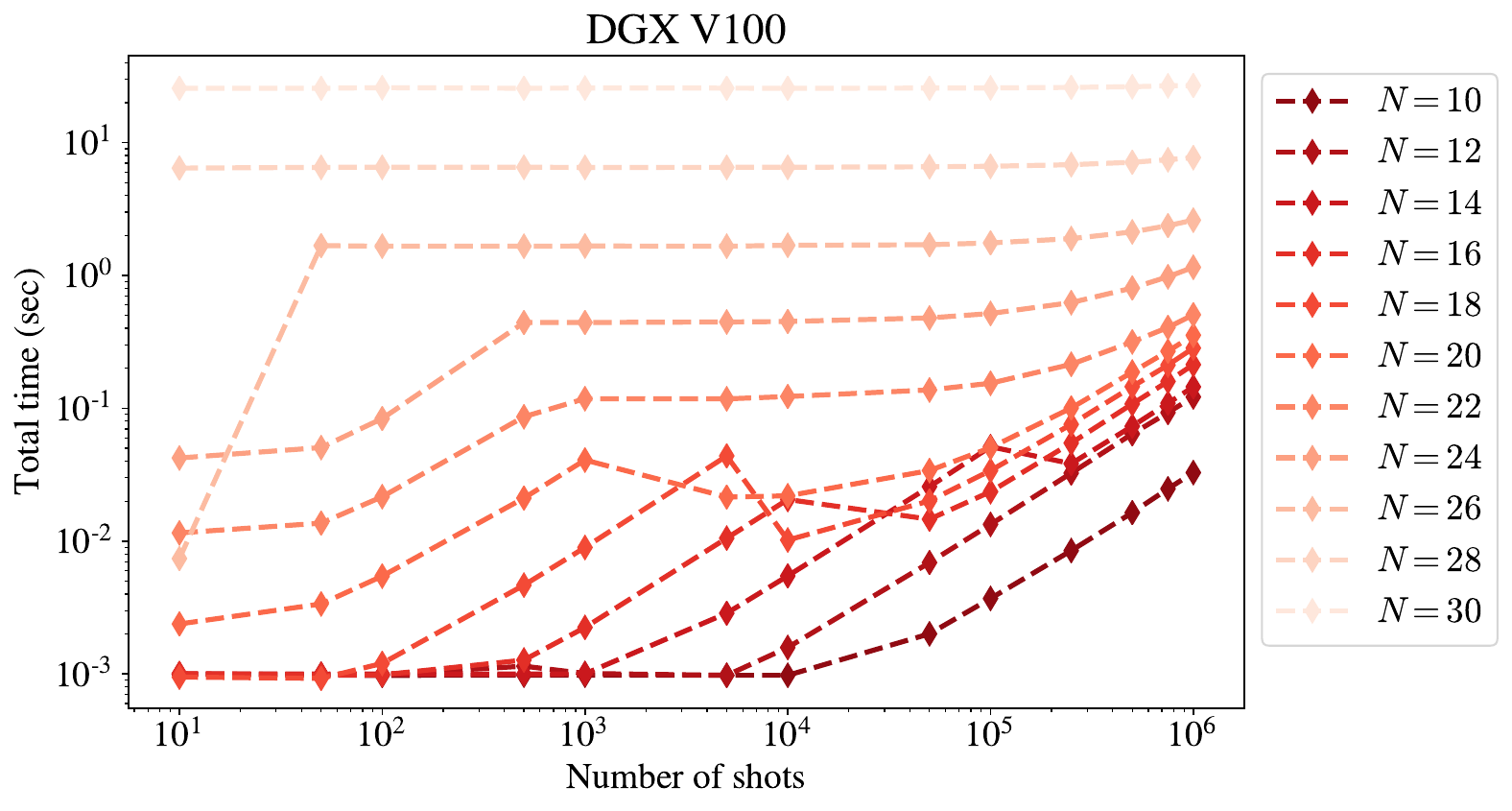}
    \caption{Example of measurement shots simulation on CPU (left) and GPU (right).}
    \label{fig:qiboshots}
\end{figure*}

\subsection{Measurement simulation}
{\tt Qibo} simulates quantum measurements using its standard dense state vector
simulator, followed by sampling from the distribution corresponding to the final
state vector. Since the dense state vector is used instead of repeated circuit
executions, measurement time does not have a strong dependence on the number of
shots. This is demonstrated for different qubit numbers $N$ in
Fig.~\ref{fig:qiboshots}. The plots contain only the time required for sampling
as the state vector simulation time (time required to apply gates) has been
subtracted from the total execution time. The circuit used in this benchmark
consists of a layer of H gates applied to every qubit followed by a measurement
of all qubits.

When a GPU is used for a circuit that contains measurements, it will likely run out of memory during the sampling procedure if the number of shots
is sufficiently high. In such a case, {\tt Qibo} will automatically fall back
to CPU to complete the execution. This is not particularly costly in terms of performance as the computationally heavy part is the evolution of the state vector, which happens on GPU, and not the sampling procedure. The oscillations
that appear in the GPU part of Fig.~\ref{fig:qiboshots} are due to this fallback
mechanism. This is implemented using an exception on TensorFlow's out-of-memory
error, and as a result, the procedure of falling back to CPU is slower than
explicitly executing sampling on CPU.

\subsection{Simulation precision}
\begin{figure}[t]
    \centering
    \includegraphics[width=0.4\textwidth]{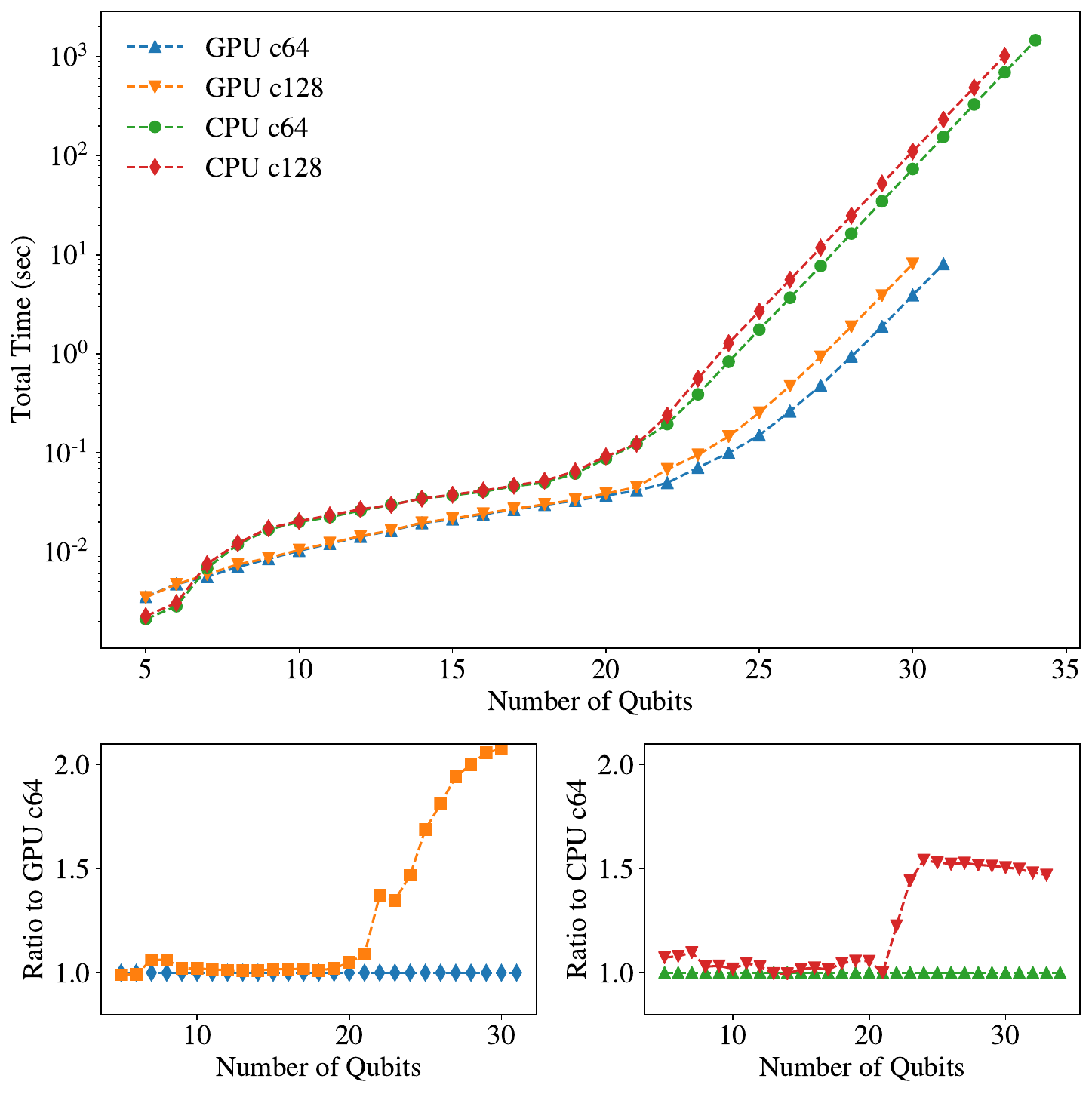}
    \caption{Comparison of simulation time when using single ({\tt complex64})
    and double ({\tt complex128}) precision on GPU and multi-threading
    (40 threads) CPU.}
    \label{fig:qiboc64vsc128}
\end{figure}

{\tt Qibo} allows the user to easily switch between single ({\tt complex64})
and double ({\tt complex128}) precision with the {\tt qibo.set\_precision()}
method. In this section we compare the performance of both precisions. The results
are plotted in Fig.~\ref{fig:qiboc64vsc128}. We find that as the number of
qubits grows using single precision is $\sim 2$ times faster on GPU and
$\sim 1.5$ faster on CPU.

\subsection{Adiabatic time evolution}\label{sec:adevbenchmarks}
\begin{figure}
    \centering
    \includegraphics[width=0.45\textwidth]{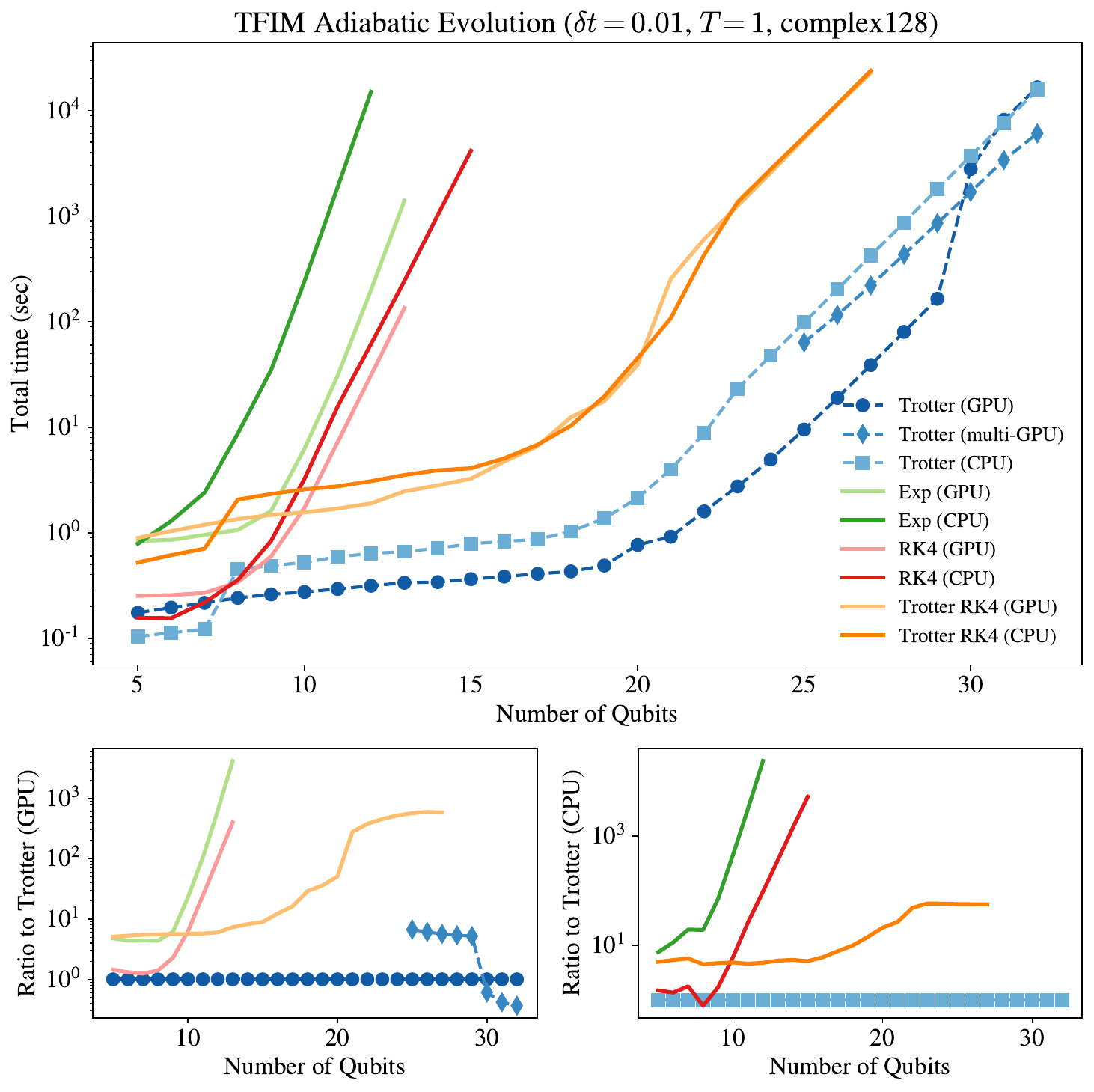}
    \caption{Adiabatic time evolution simulation performance comparison in
    double precision. Large plots show total simulation time as a function
    of qubit number. Smaller plots show the ratio of this time to the
    corresponding Trotter evolution run for GPU (left) and CPU (right).}
    \label{fig:adevbenchmarks}
\end{figure}
\begin{figure}[h!]
    \centering
    \includegraphics[width=0.425\textwidth]{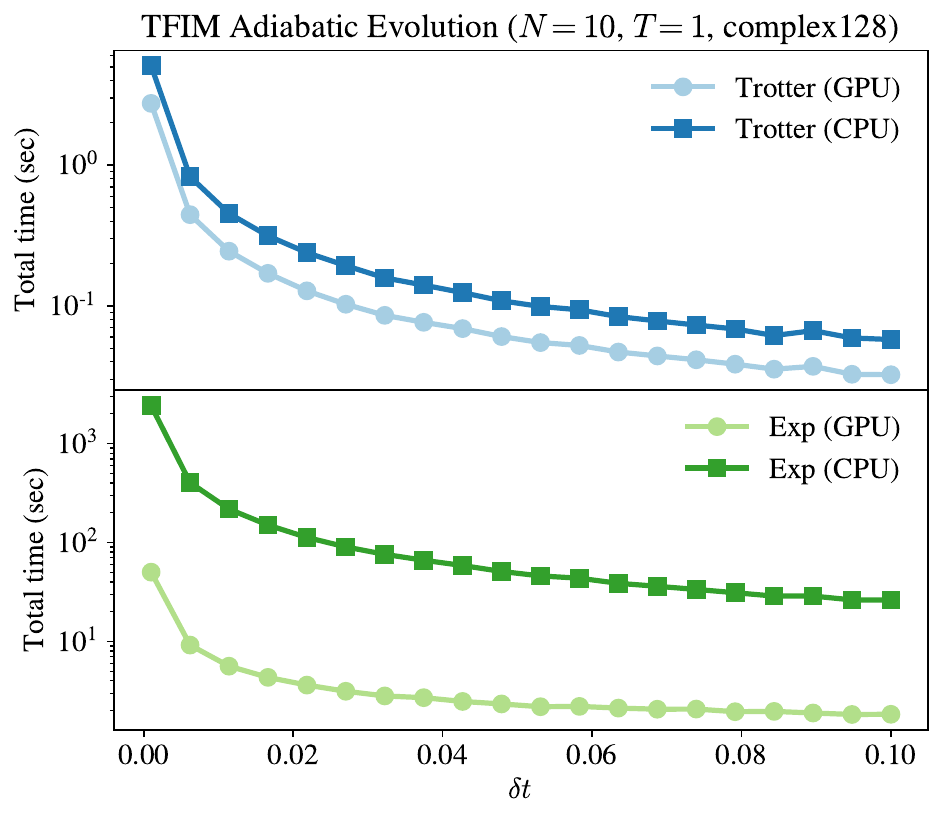}
    \caption{Total execution time for the adiabatic evolution of $N=10$ for
             total time $T=1$ as a function of the time step $\delta t$.}
    \label{fig:adevbenchmarksdt}
\end{figure}
\begin{figure}[h!]
    \centering
    \includegraphics[width=0.45\textwidth]{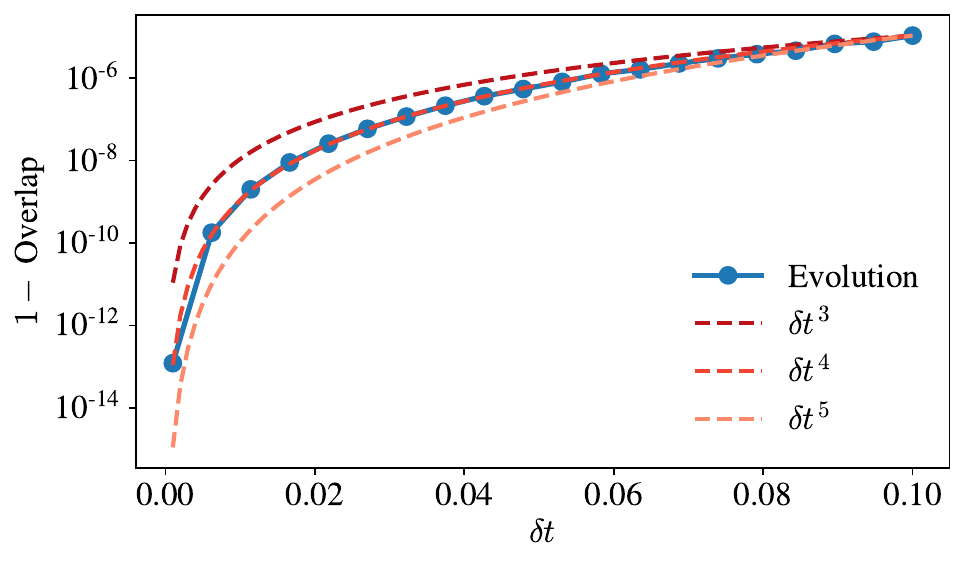}
    \caption{Overlap between the final state obtained using the Trotter decomposition
             and the full exponentiation of the Hamiltonian. The problem considered is
             the adiabatic evolution of the TFIM Hamiltonian (Eq.~(\ref{eq:tfimhamiltonian}))
             for total time $T=1$, linear scheduling and $N=10$ qubits.}
    \label{fig:trottererror}
\end{figure}

We use {\tt Qibo} to simulate the adiabatic evolution with the Hamiltonians
defined in Eq.~(\ref{eq:tfimhamiltonian}) and linear scaling $s(t) = t$. We
simulate for a total time of $T=1$ using double precision.

The total simulation time is shown as a function of the number of qubits
in Fig.~\ref{fig:adevbenchmarks} for execution on CPU (40 threads) and GPU.
As expected, using the Trotter decomposition is several orders of magnitude
faster than methods that use the full $2^N\times 2^N$ Hamiltonian matrix. It
also requires less memory allowing the simulation of larger qubit numbers.
Note that using Runge-Kutta methods with a Trotter Hamiltonian requires
less memory because it calculates the dot products between the Hamiltonian
and the state term by term, instead of constructing the full matrix.

Similar to circuit simulation, GPU is typically faster than CPU for all solvers.
Note that similarly to the QFT benchmarks shown in Fig.~\ref{fig:qft} the last
few points of the ``Trotter (GPU)'' line correspond to re-using the same device
using the distributed scheme leading to a different scaling in the execution
time.
Runge-Kutta solvers exploit parallelization techniques less than other methods and, as a result, have the smallest speed-up from using a GPU.
In all cases, the time direction has to be treated sequentially, while
the matrix multiplications at a given time step can be computed in parallel, making the GPU more useful as the number of qubits increases.
\begin{figure*}[t]
    \centering
    \includegraphics[width=0.85\textwidth]{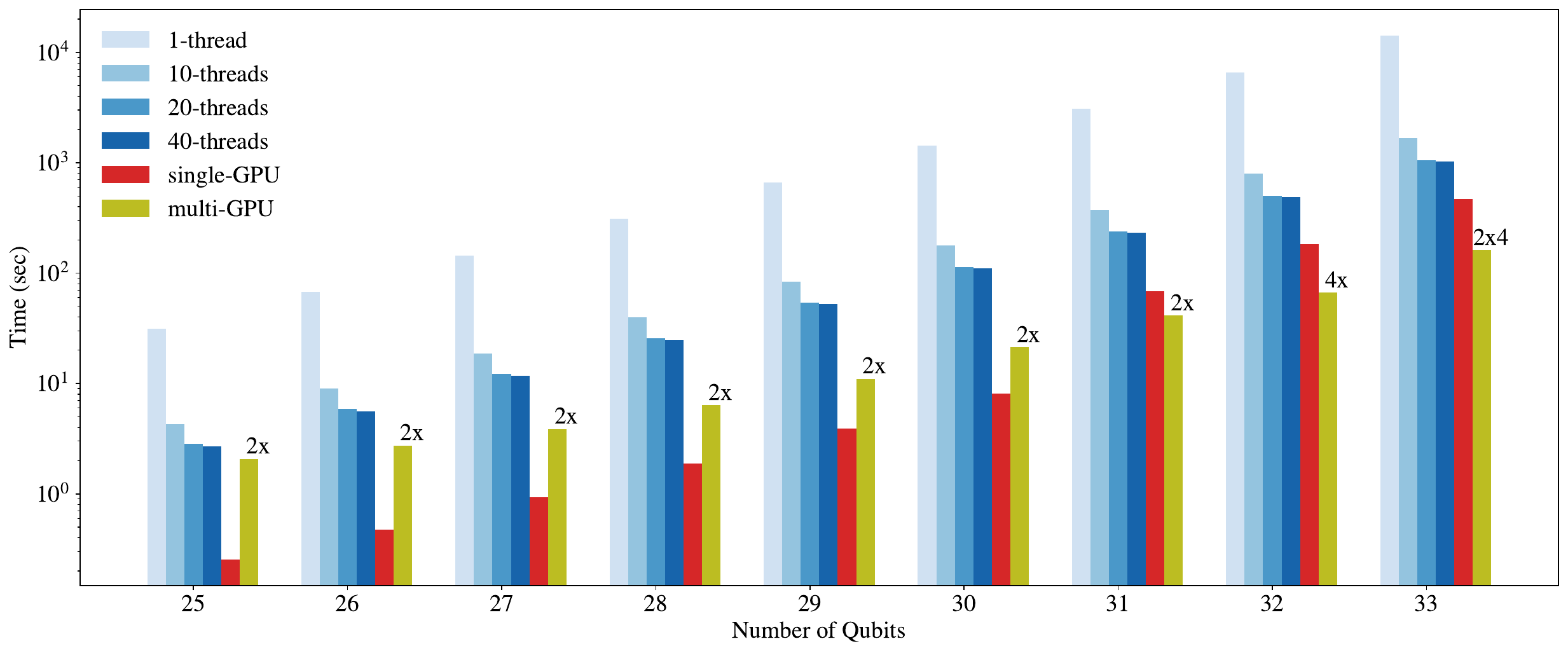}
    \caption{Comparison of {\tt Qibo} performance for QFT on multiple hardware
    configurations. For the multi-GPU setup we include a label on top of
    each histogram bar summarizing the effective number of NVIDIA V100 cards
    used during the benchmark.}
    \label{fig:configurations}
\end{figure*}
\begin{figure}[t]
    \centering
    \includegraphics[width=0.4\textwidth]{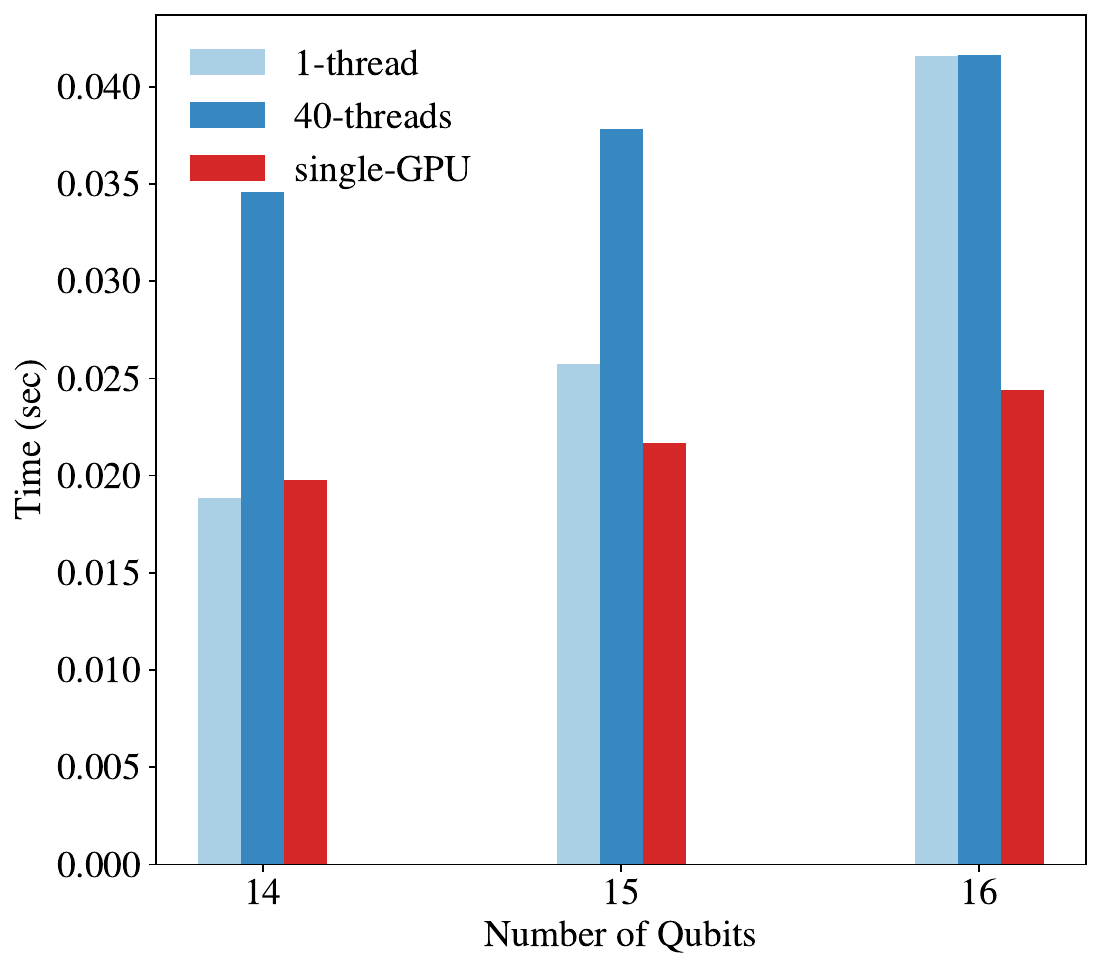}
    \caption{Comparison of {\tt Qibo} performance for small QFT circuits on
    single thread CPU, multi-threading CPU and GPU. Single thread CPU is the
    optimal choice for up to 15 qubits.}
    \label{fig:smallconfigurations}
\end{figure}

In Fig.~\ref{fig:adevbenchmarksdt} we plot the total execution time as a
function of the time step $\delta t$ used to discretize time. The exponential
solver is used with and without Trotter decomposition.
In Fig.~\ref{fig:trottererror} we calculate the underlying errors of the
Trotter decomposition. The error is quantified using the overlap between the
final state obtained using the Trotter decomposition and the full exponential
time evolution operator, with the latter considered to be exact.
We find that as we decrease the time step $\delta t$ the overlap approaches
unity as $\delta t^4$, but execution time increases as expected.
The $\delta t^n$ lines for $n\in \{3, 4, 5\}$ in Fig.~\ref{fig:trottererror}
correspond to curves defined by $y = y_0 \left (\frac{x}{x_0} \right )^n$ where
$(x_0, y_0)$ is the rightmost point (for $\delta t=0.1$) in the ``Evolution''
curve. These are plotted to demonstrate the scaling of evolution error
as $\delta t\rightarrow 0$.

\subsection{Hardware device selection}\label{sec:configurations}
\begin{table}[t]
\centering
\begin{tabular}{|l|c|c|c|}
\hline
Number of qubits & 0-15 & 15-30 & $>30$ \\ \hline \hline
CPU single thread & $\star \star \star$ & $\star $ & $\star $\\ \hline
CPU multi-threading & $\star $ & $\star \star$ & $\star \star$ \\ \hline
single GPU & $\star $ & $\star \star \star $ & $\star \star$ \\ \hline
multi-GPU & - & - & $\star \star \star$ \\ \hline
\end{tabular}
\caption{Heuristic rules for optimal device selection depending on the number
of qubits in the circuit. More stars means a shorter execution time is expected.
We stress that these general rules may not be valid on every case as the optimal
configuration depends on many factors such as the exact circuit structure and
hardware specifications (CPU and GPU speed and memory).}
\label{tab:deviceselection}
\end{table}

A core point in {\tt Qibo} is the support of different hardware configurations
despite its simple installation procedure. The user can easily switch between
CPU and GPU using the {\tt qibo.set\_device()} method. A question that arises
is how to determine the optimal device configuration for the circuit one would
like to simulate. While the answer to this question depends both on the circuit
specifics (number of qubits, number of gates, number of re-executions) and the
exact hardware specifications (CPU or GPU model and available memory), in this
section we provide a basic comparison using the DGX station. The circuit used
is the QFT in double precision, and the results of this section are summarized in
Table~\ref{tab:deviceselection}, which provides some heuristic rules for optimal
device selection according to the number of qubits.

In Fig.~\ref{fig:configurations} we plot the total simulation time using four
different CPU thread configurations and two GPU configurations. The number of
threads in the CPU runs was selected using {\tt taskset}. For large numbers of
qubits we observe that using multiple threads reduces simulation time by an
order of magnitude compared to single-thread. However, performance plateaus are reached at
20-threads. Switching to a GPU whenever the full state vector fits in its memory
provides an additional 10x speed-up making it the optimal device choice for
circuits containing 15 to 30 qubits.

The situation is less clear for circuits with more than 30 qubits, when the
full state vector does not fit in a single GPU memory. {\tt Qibo} provides
three alternative solutions for such situation: multi-threading CPU, re-using a
single GPU for multiple state pieces (mimicking distributed computation) or
using multiple GPUs if available. In terms of memory, all these approaches
are limited by the total memory available for the system's CPU. Using multiple
GPUs is always more efficient than re-using a single GPU as it allows us to
parallelize the calculations on different state pieces.

The comparison between multi-GPU and CPU generally depends on the structure of
the circuit. QFT is an example of a circuit that does not require any additional
SWAP gates between global and local qubits, making it a good case for a
distributed run. This property is not true for all circuits, and therefore we
would see a smaller difference between CPU and multi-GPU configurations for other
circuits. Regardless, multi-GPU configurations or even re-using a single GPU
are expected to be useful for the regime of qubit numbers between 30 and 35.

In Fig.~\ref{fig:smallconfigurations} we repeat the hardware comparison for
smaller qubit numbers. We find that single thread CPU is the optimal choice for
up to 15 qubits, while the GPU will start giving an advantage beyond this point.

\section{Applications}
\label{sec:applications}

The current {\tt Qibo 0.1.0} contains pre-coded examples of quantum algorithms
applied to specific problems. The subsections that follow provide an outline of
these applications. For more details on each application we refer to our
documentation~\cite{qibo_docs} and we note that all the code is available at the
{\tt Qibo} repository.

It is worth emphasizing that, apart from the application examples that follow,
{\tt Qibo} provides several application-specific models, in addition to the
standard {\tt models.Cir-\\cuit} that can be used for circuit simulation.
These models are outlined in Table~\ref{tab:qibomodels} and some are used
in the applications that follow.

\subsection{Variational Quantum Eigensolver}
The Variational Quantum Eigensolver (VQE)~\cite{vqe} is a common technique
for finding ground states of Hamiltonians in the context of quantum
computation. As noted in Table~\ref{tab:qibomodels}, {\tt Qibo} provides
a VQE model that allows optimization of the variational parameters.

In this example, we provide an extension of the VQE algorithm
that may be used to improve optimization and is known as the adiabatically
assisted VQE (AAVQE)~\cite{aavqe}.
{\tt Qibo} provides a pre-coded implementation of the AAVQE for finding
the ground state of the transverse field Ising Hamiltonian defined in
Eq.~(\ref{eq:tfimhamiltonian}) and can be executed for any number of qubits,
variational circuit layers and number of adiabatic steps specified by the user.
Particularly, the example may be used to explore how the accuracy of
the VQE ansatz scales with the underlying circuit depth, as presented
in~\cite{scalingvar}.

The code below demonstrates how the AAVQE can be implemented in {\tt Qibo}.
\begin{minted}{python}
# loop over the adiabatic steps
for t in range(T_max + 1):
    # define the interpolated Hamiltonian of AAVQE
    s = t / T_max
    hamiltonian = (1-s) * h0 + s * h1
    # define a Qibo VQE model
    vqe = models.VQE(circuit, hamiltonian)
    # optimize the VQE model to find that ground state
    options = {'maxfev': maxsteps}
    energy, params = vqe.minimize(initial_params,
                                  method='Nelder-Mead',
                                  options=options)
    initial_params = params
\end{minted}
where {\tt h0} and {\tt h1} are {\tt hamiltonians.Hamiltonian} objects
representing the easy and hard Hamiltonian respectively,
{\tt circuit}  is a {\tt models.Circuit} that implements the VQE ansatz,
{\tt initial\_params (np.ndarray)} is the  initial guess for the
variational parameters, {\tt T\_max (int)} the number of adiabatic steps
and {\tt maxsteps (int)} the maximum number of optimization steps.

\subsection{Grover's search for 3SAT}\label{sec:grover3sat}
Grover's algorithm~\cite{grover1996fast} is a quantum algorithm
for searching databases and an example where a quantum computer provides a
quadratic advantage $\sqrt{2 ^N}$, where $N$ is the number of qubits,
over a classical one for the same problem.
In this example, Grover's algorithm is used to solve exact cover
instances of a 3SAT problem, which is classified as NP-Complete \cite{karp1975computational}.

In terms of implementation, Grover's algorithm can be simulated by
defining a {\tt Qibo} circuit that contains the required gates, which
implement the \textit{oracle} and the \textit{diffusion} transformation.
The pre-coded example comes with several instances of the exact cover
problem from 4 up to 30 bits, where the solution is known, so that the user
can verify that the algorithm has been successful, and that the solution is indeed the measured bitstring. Moreover, the user may execute
the algorithm to newly created instances, with a known or unknown solution.

\subsection{Grover's search for hash functions}
A second application of Grover's algorithm~\cite{grover1996fast}
implemented in {\tt Qibo} is on the task of finding the preimages
of a hash function based on the ChaCha permutation~\cite{chachaclassical}.
The example is based on Ref.~\cite{hash-preimage}.

In this case, the example takes as input a {\tt hash} integer of 8
or fewer bits and finds the corresponding preimage, that is the number
that maps to the given {\tt hash} when applying the permutation. If
the number of {\tt collisions} is known for the given {\tt hash}
then the algorithm finds all the possible solutions.
Here {\tt collisions} refers to the number of solutions.
If the number of {\tt collisions} is not given then the algorithm
finds one solution using an iterative procedure~\cite{iterativegrover}.

\subsection{Quantum classifier}
This example provides a variational quantum algorithm for classifying
classical data~\cite{qembedings}. The optimal values for the variational
parameters are found via supervised training, minimizing a local loss
given by the quadratic deviation of the classifier's predictions from
the actual labels of the examples in the training set.

The pre-coded {\tt Qibo} example applies the classifier on the iris
data set~\cite{irisdataset}. The user has the option to train the
circuit from scratch or use several pre-trained configurations to
make predictions and calculate the classification accuracy. The
user can also select the number of qubits and the circuit depth.

\subsection{Quantum classifier using data reuploading}
Similarly to the previous section, this example provides a variational
algorithm for classifying classical data using only one qubit and is
based on Ref.~\cite{datareuploading}.
The main idea is \textit{reuploading}, that is creating a single qubit
circuit where several different unitary gates are applied and each
gate depends on the point that is classified and a set of variational
parameters that are optimized through a learning procedure.

The pre-coded {\tt Qibo} example applies such classifier on various
two-dimensional classical datasets. The user can choose between
training the classifier from scratch (optimizing the variational
parameters) or using the provided pre-trained models. The code can
be used to measure the accuracy of the classifier in each classification
task and also provides plots with labeled points, using different colors
for each class. Plots are provided in the two-dimensional plane but also
on the Bloch sphere.

\subsection{Quantum autoencoder for data compression}
The task of an autoencoder is to map an input density matrix $\mathbf{x}$ to
a lower-dimensional Hilbert space $\mathbf{y}$, such that $\mathbf{x}$ can
be recovered from $\mathbf{y}$. The quantum autoencoder that is
implemented in {\tt Qibo} is based on~\cite{qautoencoder} and is
used to encode the ground states of the transverse field Ising model
(Eq.~(\ref{eq:tfimhamiltonian})) for various $h$-field values.

The code below demonstrates how the autoencoder optimization can be
simulated using {\tt Qibo}
\begin{minted}{python}
def cost(params):
    """Calculates loss for specific values of parameters."""
    circuit.set_parameters(params)
    for state in ising_groundstates:
        final_state = circuit(np.copy(state))
        cost += encoder.expectation(final_state).numpy().real

options = {'maxiter': 2.0e3, 'maxfun': 2.0e3}
result = minimize(cost, initial_params,
                  method='L-BFGS-B',
                  options=options)
\end{minted}
where {\tt circuit} is a {\tt Qibo} circuit that implements the
variational ansatz, {\tt ising\_groundstates} is a list of the
states to be encoded and {\tt encoder} is a {\tt Hamiltonian}
object for $-\sum _{i=1}^NZ_i$ properly rescaled.

\subsection{Quantum singular value decomposer}
The quantum singular value decomposer (QSVD)~\cite{qsvd} refers
to a circuit that produces the Schmidt coefficients of a pure bipartite
quantum state. This is implemented as follows: Two {\tt Qibo}
variational circuits are defined and measured, one acting on each
part of the bipartite measurements. The variational parameters are
tuned to minimize a loss that depends on the Hamming distance
of the two measured bitstrings. The user may attempt this optimization
using the pre-coded example for random initial bipartite states with
an arbitrary number of qubits and partition sizes.

\subsection{Tangle of three-qubit states}
This example provides a variational strategy to compute the tangle
of an unknown three-qubit state, given many copies of it. It is based
on the result that local unitary operations cannot affect the entanglement
of any quantum state and follows Refs.~\cite{tangle, tangle2}.
An unknown three-qubit state is received, and one unitary gate is applied
on every qubit. The exact gates are obtained such that they minimize the
number of outcomes of the states $|001\rangle $, $|010\rangle $ and
$|011\rangle $. The code can be used to simulate both noiseless and
noisy circuits.

\subsection{Unary approach to option pricing}
This is an example application of quantum computing in finance and provides
a new strategy to compute the expected payoff of a (European) option,
based on Ref.~\cite{unary}. The main feature of this procedure is to use
the unary encoding of prices, that is, to work in the subspace of the Hilbert
space spanned by computational-basis states where only one qubit is in the $|1\rangle $ state.
This allows for a simplification of the circuit and resilience against
errors, making the algorithm more suitable for NISQ devices.

The pre-coded example takes as input the asset parameters (initial price,
strike price, volatility, interest rate and maturity date) and
the quantum simulation parameters (number of qubits, number of measurement
shots and number of applications of the amplitude estimation algorithm
(see~\cite{unary})) and calculates the expected payoff using the
quantum algorithm. The result is compared with the exact value of the
expected payoff. Furthermore, the code plots a histogram of the quantum
estimation for the option price probability distribution and a plot
of the amplitude estimation measurement results as a function of iterations.

\subsection{Adiabatic evolution}
As noted in Table~\ref{tab:qibomodels}, {\tt Qibo} provides models
for simulating the unitary time evolution of quantum states, with
a specific application on adiabatic evolution. As examples, we
provide scripts that apply these models in various physics applications.

The first example simulates the adiabatic evolution of an Ising Hamiltonian
(Eq.~(\ref{eq:tfimhamiltonian})) for $h=1$ using a linear
scaling function $s(t) = t / T$, where $T$ is the total evolution time.
Executing this example shows plots with the dynamics of energy (expectation
value of the Ising Hamiltonian) and the overlap between the evolved state
and the exact Ising ground state. Using these plots, we verify that the
evolution converges to the exact ground state if sufficient evolution time $T$
is used.

In addition, we provide the possibility to optimize the scheduling
function $s(t)$ and final time $T$ so that the actual ground state is
reached in a shorter time. The free parameters of $s(t)$ and $T$ are
optimized so that the energy of the final state of the evolution is
minimized. In our example, we use a polynomial ansatz for $s(t)$ where
the coefficients are the free parameters that are optimized.

\subsection{Adiabatic evolution for 3SAT}
Adiabatic evolution can also be applied to optimization problems outside physics.
In this example, we provide an application for solving exact cover instances
of a 3SAT problem, which is classified as an NP-Complete problem \cite{karp1975computational}.
The same problem was solved in Sec.~\ref{sec:grover3sat} using Grover's
algorithm in the circuit based paradigm of quantum computation,
while in this example we demonstrate that a quantum annealing approach
is also possible~\cite{adiabatic3sat}.

Similarly to Sec.~\ref{sec:grover3sat}, the user may use one of the
provided instances of the exact cover problem or add a custom one.
The pre-coded example accepts the instance and the evolution parameters
(total time $T$, discretization time step $\delta t$, and method of
integration) and computes the solution and the probability that this
is measured. The Trotter decomposition may be used for a more efficient
evolution simulation.
Additionally, for sufficiently small systems (due to memory
constraints), it is possible to calculate and plot the gap of the
adiabatic Hamiltonian as a function of time.

This example uses a linear scaling function by default. It is possible
to switch to a polynomial scaling function and also optimize the
underlying coefficient so that the solution is found in smaller total
time $T$. Performing such optimization, we find that a scaling function
that is ``slower" at times where the gap is small is preferred over
the default linear scheduling.

\section{Outlook}
\label{sec:outlook}

{\tt Qibo} provides a new interface for quantum simulation research by granting
users and researchers the ability to implement quantum algorithms with
simplicity. The user is allowed to simulate circuits and adiabatic evolution on
different hardware platforms without having to know about the technicalities or
the difficulties of their implementation on data placement, multi-threading
systems and memory management that GPU and multi-GPUs computing require.

In this first release, {\tt Qibo} includes a high-performance framework for
quantum circuit simulations and adiabatic evolution using linear algebra
techniques in combination with hardware acceleration strategies. For the time
being, the code is targeted to run on single node devices with single or multiple
GPU cards and sufficient RAM in order to perform simulations with an
acceptable number of qubits and computational time.

The roadmap for future releases is organized in two directions. The first
direction is focused on Physics and new algorithms for specific applications,
in order to extend the code base set of algorithms for quantum calculations.
Some imminent examples are the implementation of noise density matrices for
custom operators and noise simulation without density matrices. The second
direction is based on the technical perspective. We plan to extend the current
distributed computation model to support multi-node devices through the {\tt
OpenMPI}~\cite{OpenMPI1,OpenMPI2} interface.

\begin{figure}
    \includegraphics[width=0.45\textwidth]{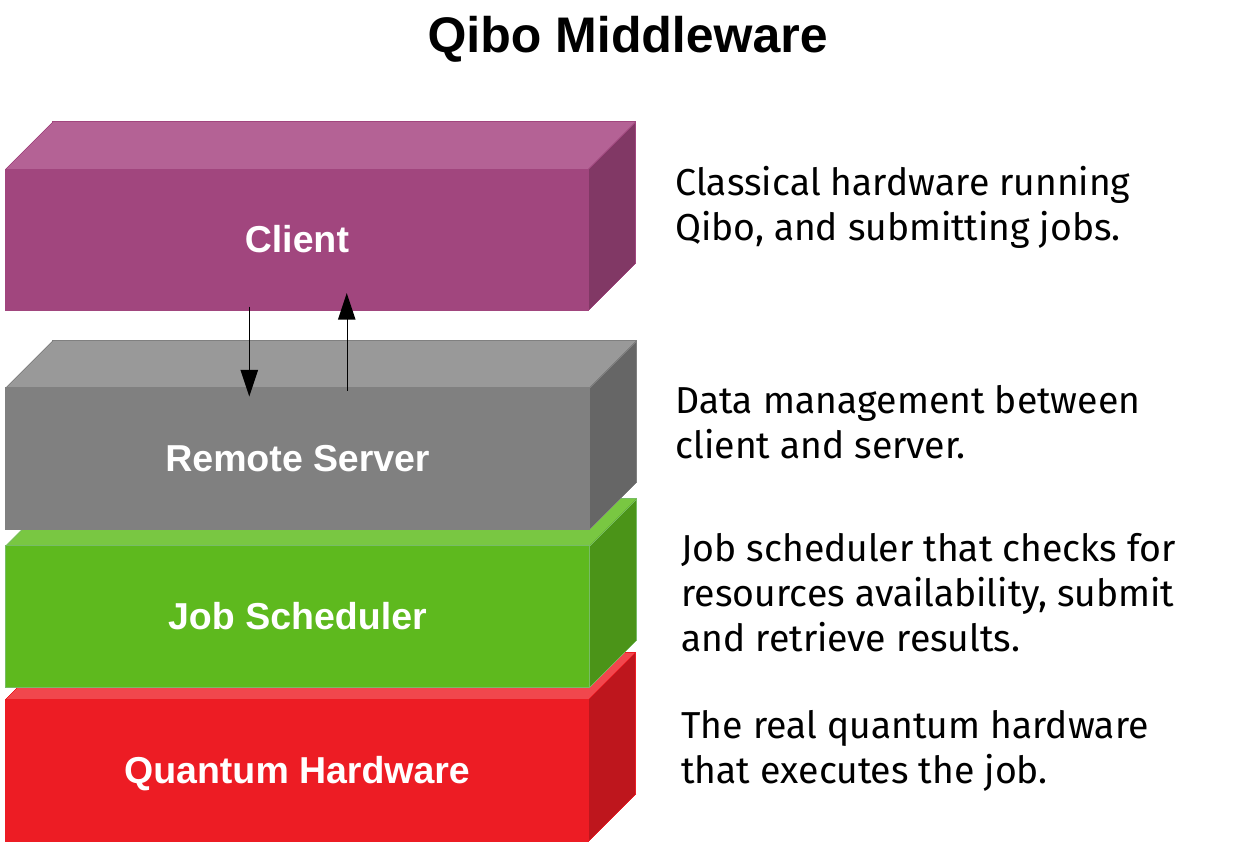}
    \caption{Schematic view of the {\tt Qibo} middleware design.}
    \label{fig:middleware}
\end{figure}

Furthermore, in Fig.~\ref{fig:middleware} we show schematically how the {\tt
Qibo} framework will be integrated with the middleware of the new quantum
hardware developed by~\cite{TII,QQT}. The middleware infrastructure will provide
the possibility to submit quantum calculations, defined with the {\tt Qibo} API,
to the quantum hardware through a server scheduling and queue service that
provide the possibility to submit and retrieve results from the quantum
computer laboratory. This development is particularly important in order to
achieve the evaluation of quantum circuits, adiabatic evolution, and hybrid
computations on the real quantum hardware.

\section*{Acknowledgements}

The {\tt Qibo} framework is supported by the Quantum Research Centre at the
Technology Innovation Institute in the United Arab Emirates~\cite{TII} and the
Qilimanjaro Quantum Tech in Spain~\cite{QQT}. This work is supported by project
QuantumCAT (ref.~001-P-001644), co-funded by the Generalitat de Catalunya and
the European Union Regional Development Fund within the ERDF Operational
Program of Catalunya, and the European Union's Horizon 2020 research and innovation
programme under grant agreement No 951911 (AI4Media).

\bibliographystyle{elsarticle-num}
\bibliography{qibo.bib}

\end{document}